\documentclass{emulateapj}

\usepackage{graphicx}
\usepackage{amssymb}
\usepackage{longtable}
\usepackage{epsf}
\usepackage{apjfonts}

\slugcomment{Accepted in the Astrophysical Journal}

\shorttitle{The progenitors of local ultra-massive galaxies since $z=3$}
\shortauthors{Marchesini et al.}

\begin{document}

\title{The Progenitors of Local Ultra-massive Galaxies Across Cosmic Time: \\
from Dusty Star-bursting to Quiescent Stellar Populations}

\author{Danilo Marchesini\altaffilmark{1}, Adam Muzzin\altaffilmark{2}, 
Mauro Stefanon\altaffilmark{3}, Marijn Franx\altaffilmark{2}, 
Gabriel G. Brammer\altaffilmark{4}, Cemile Z. Marsan \altaffilmark{1}, 
Benedetta Vulcani \altaffilmark{5}, J.~P.~U.~Fynbo\altaffilmark{6}, 
Bo Milvang-Jensen\altaffilmark{6}, James~S.~Dunlop\altaffilmark{7}, 
Fernando~Buitrago\altaffilmark{7}}

\altaffiltext{1}{Department of Physics and Astronomy, Tufts University, 
Medford, MA 02155, USA}
\altaffiltext{2}{Leiden Observatory, Leiden University, PO Box 9513, 
NL-2300 RA Leiden, The Netherlands}
\altaffiltext{3}{Physics and Astronomy Department, University of Missouri, 
Columbia, MO 65211, USA}
\altaffiltext{4}{Space Telescope Science Institute, 3700 San Martin Drive, 
Baltimore, MD 21218, USA} 
\altaffiltext{5}{Kavli Institute for the Physics and Mathematics of the 
Universe (WPI), Todai Institutes for Advanced Study, University of Tokyo, 
Kashiwa 277-8582, Japan} 
\altaffiltext{6}{Dark Cosmology Centre, Niels Bohr Institute, University of 
Copenhagen, Juliane Maries Vej 30, DK-2100 Copenhagen, Denmark} 
\altaffiltext{7}{SUPA, Institute for Astronomy, University of Edinburgh, 
Royal Observatory, Edinburgh EH9 3HJ, UK}

\begin{abstract}
Using the UltraVISTA catalogs, we investigate the evolution in the 11.4~Gyr 
since $z=3$ of the progenitors of local ultra-massive galaxies 
($\log{(M_{\rm star}/M_{\odot})}\approx11.8$; UMGs), providing a complete 
and consistent picture of how the most massive galaxies at $z=0$ have 
assembled. By selecting the progenitors with a semi-empirical approach using 
abundance matching, we infer a growth in stellar mass of 
0.56$^{+0.35}_{-0.25}$ dex, 0.45$^{+0.16}_{-0.20}$~dex, and 0.27$^{+0.08}_{-0.12}$ 
dex from $z=3$, $z=2$, and $z=1$, respectively, to $z=0$. At $z<1$, the 
progenitors of UMGs constitute a homogeneous population of only quiescent 
galaxies with old stellar populations. At $z>1$, the contribution from 
star-forming galaxies progressively increases, with the progenitors at $2<z<3$ 
being dominated by massive ($M_{\rm star} \approx 2 \times 10^{11}$M$_{\odot}$), 
dusty ($A_{\rm V}\sim$1--2.2 mag), star-forming (SFR$\sim$100--400~M$_{\odot}$ 
yr$^{-1}$) galaxies with a large range in stellar ages. At $z=2.75$, 
$\sim$15\% of the progenitors are quiescent, with properties typical of 
post-starburst galaxies with little dust extinction and strong Balmer break, 
and showing a large scatter in color. Our findings indicate that at least half 
of the stellar content of local UMGs was assembled at $z>1$, whereas the 
remaining was assembled via merging from $z\sim 1$ to the present. Most of the 
quenching of the star-forming progenitors happened between $z=2.75$ and 
$z=1.25$, in good agreement with the typical formation redshift and scatter in 
age of $z=0$ UMGs as derived from their fossil records. The progenitors of 
local UMGs, including the star-forming ones, never lived on the blue cloud 
since $z=3$. We propose an alternative path for the formation of local UMGs 
that refines previously proposed pictures and that is fully consistent with 
our findings. 
\end{abstract}

\keywords{galaxies: evolution --- galaxies: formation --- 
galaxies: fundamental parameters --- galaxies: high-redshift --- 
galaxies: luminosity function, mass function --- 
galaxies: stellar content}


\section{Introduction}

One of the most controversial questions regarding the formation and evolution 
of galaxies is when and how today's most massive galaxies formed. In the 
standard paradigm of structure formation, dark matter halos build in a 
hierarchical fashion through the dissipationless mechanism of gravitational 
instability, while galaxies form inside these structures following the 
radiative cooling of baryons, with stars forming out of gas that cools within 
the halos. In this picture, the most massive galaxies are naturally at the 
endpoint of the hierarchical merging process, and offer an opportunity to 
constrain models of galaxy formation and evolution, since their predicted 
galaxy properties are sensitive to various model assumptions, such as gas 
cooling, star formation, stellar and AGN feedback processes, chemical 
evolution, and mergers (e.g., \citealt{baugh06}; \citealt{delucia06}; 
\citealt{henriques12}; and references therein).

In the local universe ($z\sim0$), the most massive galaxies (i.e., stellar mass 
$M_{\rm star}>5\times10^{11}$~M$_{\odot}$) constitute a homogeneous population. 
These local ultra-massive galaxies tend to live in high-density 
environments (typically rich groups or clusters of galaxies; e.g., 
\citealt{blanton09} for a review) and are characterized by early-type 
morphologies, red rest-frame optical colors, extremely quiescent stellar 
populations, old mean stellar ages, significant enhancements of the $\alpha/Fe$ 
element ratio, and little dust attenuation, and populate the extreme massive 
end of the very tight red sequence (e.g., \citealt{bower92}; 
\citealt{kauffmann03}; \citealt{kauffmann04}; \citealt{gallazzi05}; 
\citealt{nelan05}; \citealt{thomas05}; \citealt{baldry06}; 
\citealt{gallazzi06}; \citealt{thomas10}). In particular, two decades of 
archeological studies (i.e., the detailed investigation of the stellar 
populations in local galaxies to constrain their formations and evolutions 
with cosmic time) have robustly shown that the more massive the galaxy is 
today, the earlier its star formation must have started and more promptly must 
have subsided, in an apparently ``anti-hierarchical'' fashion of the star 
formation. Quantitatively, most of the stars in local ultra-massive galaxies 
must have formed in the first $\sim3$~Gyr of cosmic history (i.e., $z>2$) 
through short ($<$1~Gyr, as low as $\sim0.2$~Gyr for the most massive 
galaxies), hence intense, bursts of star formation (e.g., \citealt{renzini06}; 
\citealt{vandokkum07}; \citealt{thomas10}). 

Increasingly more sophisticated models of galaxy formation and evolution have 
shown that the apparent down-sizing in the star formation of ultra-massive 
galaxies is not in contradiction with the hierarchical paradigm. Recent models 
do predict `anti-hierarchical' star-formation histories for the most massive 
galaxies in a $\Lambda$CDM universe (with most stars formed by $z\sim3$), 
whereas the assembly of these galaxies is indeed hierarchical, with $\sim$80\% 
of the final mass typically locked up in a single galaxy only at $z<0.4$ and 
with a number of effective progenitors as high as $\sim$5 for the most massive 
galaxies (e.g., \citealt{delucia06}; \citealt{delucia07}). 

These predictions are qualitatively in agreement with the nearly constant 
rest-frame optical luminosity density of red-sequence galaxies since $z\sim1$ 
found by the COMBO-17 and the DEEP2 surveys (\citealt{bell04}; 
\citealt{faber07}). This result in fact implies that the stellar mass density 
in red-sequence galaxies has roughly doubled since $z\sim1$ driven by a 
combination of galaxy merging and truncation of star formation of the blue 
star-forming population. In particular, \citet{faber07} have used these 
results in support of a scenario for the formation of massive red-sequence 
(spheroidal) galaxies involving early mass assembly and star formation (with 
such progenitors living on the blue cloud), followed by quenching (which 
migrates the progenitors onto the red sequence) and further dry merging 
(allowing for additional growth along the red sequence to explain the assembly 
of the most massive galaxies in the local universe). 

However, an independent analysis by \citet{cimatti06} of the evolution of the 
rest-frame optical luminosity function of red-sequence galaxies since 
$z\sim0.8$ from COMBO-17 and DEEP2 has shown that the massive 
($M_{\rm star}>10^{11}$~M$_{\odot}$) red-sequence galaxies have actually not 
grown in mass over the past $\sim$7 billion years, while the inferred growth 
by a factor of $\sim$2 of the stellar mass density of red-sequence galaxies 
since $z\sim0.8$ appears due to the build-up of the less-massive red-sequence 
population. In addition to the anti-hierarchical formation of the stars, this 
result implies also an anti-hierarchical assembly of the stellar content of 
the most massive galaxies, in contrast to the predictions from theoretical 
models. 

Recent measurements of the stellar mass function of galaxies out to redshift 
$z=4$ have intensified the tension between observations and theoretical 
predictions. Indeed, increasingly wider and deeper near-infrared surveys have 
provided evidence for the existence of very massive galaxies 
(i.e., $\log{(M_{\rm star}/M_{\odot})} \gtrsim 11.5$) when the universe was only 
1.5 Gyr old (i.e., $z \sim 4$), and of their number densities evolving very 
little in the following $\sim$3--4~Gyr from $z=4$ to $z\sim$1--1.5 (e.g., 
\citealt{perez08}; \citealt{marchesini09}; \citealt{marchesini10}; 
\citealt{brammer11}; \citealt{muzzin13b}; and references therein). 

While the stellar mass function of galaxies provides a mean to measure the 
abundance of a galaxy population at a given time and the overall growth as a 
function of time of its stellar content, it does not tell us how individual 
galaxies have evolved and assembled their mass. Ultimately, we would like to 
be able to connect local ultra-massive galaxies with their progenitors in the 
early universe to understand how their stellar populations have actually 
changed over cosmic time and to answer the following outstanding questions. Do 
local ultra-massive galaxies really form in short and intense bursts of star 
formation at high redshift? When do they actually stop forming stars? How does 
their stellar population age? How much mass is assembled over time? Do they 
really evolve from the blue cloud to the red sequence as suggested by, e.g., 
\citet{faber07}? 

Needless to say, answering these questions have proved difficult, as it 
requires the non-trivial task of linking galaxies and their 
descendants/progenitors through cosmic time, which in turn requires 
assumptions for how galaxies evolve. However, in recent years, several 
approaches have been developed to link galaxies across cosmic time. A popular 
approach easy to implement is to compare galaxy properties at fixed cumulative 
number density (e.g., \citealt{wake06}; \citealt{tojeiro10}; 
\citealt{brammer11}; \citealt{papovich11}; \citealt{tojeiro12}; 
\citealt{patel13}; \citealt{vandokkum13}; \citealt{leja13}; 
\citealt{muzzin13b}). While this approach has the advantage of relying solely 
on observations, it does not take into account galaxy-galaxy mergers and 
scatter in mass accretion histories, both affecting the median cumulative 
number density of a galaxy population (\citealt{leja13}; \citealt{lin13}; 
\citealt{behroozi13a}). More advanced comparisons have adopted either 
semi-analytical or semi-empirical galaxy-halo connections to deduce galaxy 
evolution from simulated dark matter merger histories (e.g., 
\citealt{conroy09}; \citealt{leitner12}; \citealt{behroozi13a}; 
\citealt{moster13}; \citealt{wang13}; \citealt{lu12}; \citealt{leja13}; 
\citealt{lin13}; \citealt{mutch13}; \citealt{lu14}; \citealt{behroozi13b}; and 
references therein). 

In this paper, we investigate the evolution with cosmic time of the 
progenitors of local ultra-massive ($M_{\rm star} \approx 6 \times 10^{11}$ 
M$_{\odot}$) galaxies from $z=3$. The progenitors are selected using both the 
fixed cumulative number density technique and a semi-empirical approach using 
abundance matching in the $\Lambda$CDM paradigm. The stellar population 
properties of the progenitors are then studied as a function of redshift. This 
analysis provides a direct test of the star-formation histories predicted from 
the archeology studies of local ultra-massive galaxies and returns a clear and 
consistent picture of their evolution in the past 11.4 billion years of cosmic 
history. 

Our paper is structured as follows: The adopted dataset and the selection of 
the sample of progenitors are described in \S\ref{sec-dataset} and 
\S\ref{sec-sample}, respectively. The evolution of the properties of the 
progenitors of local ultra-massive galaxies is presented in 
\S\ref{sec-results}, while the results are summarized and discussed in 
\S\ref{sec-discussion}. We assume $\Omega_{\rm M}=0.3$, 
$\Omega_{\rm \Lambda}=0.7$, and $H_{\rm 0}=70$~km~s$^{-1}$Mpc$^{-1}$. A 
\citet{kroupa01} initial mass function (IMF) is assumed throughout the paper. 
All magnitudes are in the AB system.


\section{The Dataset}\label{sec-dataset}

This study uses the K$_{\rm S}$-selected catalog of the COSMOS/UltraVISTA field 
from \citet{muzzin13a}. The catalog includes PSF-matched photometry in 30 
photometric bands over the wavelength range 0.15$\mu$m $\rightarrow$ 24$\mu$m 
from the available GALEX \citep{martin05}, CFHT/Subaru \citep{capak07}, 
UltraVISTA \citep{mccracken12}, and S-COSMOS \citep{sanders07} datasets. 
Sources are selected from the DR1 
UltraVISTA K$_{\rm S}$-band imaging \citep{mccracken12} which reaches a depth 
of K$_{\rm S,tot}<23.4$ at 90\% completeness. A detailed description of the 
photometric catalog construction, photometric redshift measurements, and 
stellar population properties' estimates (e.g., stellar mass) is presented in 
\citet{muzzin13a}, which accompanies the public release of all data products 
from the catalog. A brief description of the relevant aspects of the catalog 
with respect to our analysis is also provided in \citet{muzzin13b}, presenting 
the measurements of the stellar mass function of galaxies, including quiescent 
and star-forming galaxies, from $z=4$ to $z=0.2$. 

Briefly, stellar population properties were derived by fitting the observed 
spectral energy distributions (SEDs) from the {\it GALEX} UV to the 
{\it Spitzer}-IRAC 8$\mu$m photometry with \citet{bruzual03} models assuming 
exponentially-declining star-formation histories (SFHs) of the form 
SFR$\propto e^{\rm -t/\tau}$, where SFR is the star formation rate, $t$ is the 
time since the onset of star formation, and $\tau$ sets the timescale of the 
decline in the SFR, solar metallicity, a \citet{calzetti00} dust law, and a 
\citet{kroupa01} IMF. We allow $\log{(\tau/Gyr)}$ to range between 7.0 and 
10.0 Gyr, $\log{(t/Gyr)}$ between 7.0 and 10.1 Gyr, and $A_{\rm V}$ between 0 
and 4 mag. Median random errors on stellar mass, SFR, stellar age, and dust 
extinction are $^{+0.05}_{-0.09}$~dex, $\pm$0.2~dex, $^{+0.1}_{-0.2}$~dex, and 
$^{+0.3}_{-0.2}$~mag, respectively, for the quiescent progenitors, and 
$^{+0.09}_{-0.18}$~dex, $\pm$0.6~dex, $^{+0.2}_{-0.4}$~dex, and $^{+0.6}_{-0.4}$ 
mag, respectively, for the star-forming progenitors, similar to the typical 
uncertainties from low-resolution spectroscopy \citep{muzzin09}. To assess the 
systematic effects of the adopted SED-modeling assumptions, we have also 
investigated different SFHs, metallicities, and extinction curves (see 
\S\ref{s-sedmodel}).

The SFRs from SED modeling can be strongly influenced by the assumption of the 
SFH. A more robust estimate of the instantaneous SFR is obtained from the 
combination of the rest-frame UV flux ($L_{\rm 2800}$) and the rest-frame total 
infrared luminosity ($L_{\rm IR}$). To determine $L_{\rm 2800}$ we used EAZY 
\citep{brammer08} to integrate the best-fit template over the wavelength range 
2600-2950\AA. The inclusion of the {\it GALEX} photometry in the UltraVISTA 
multiwavelength catalogs of \citet{muzzin13a} guarantees that $L_{\rm 2800}$ is 
constrained by the data over the full targeted redshift range. The approach 
presented in \citet{wuyts08} was followed to determine $L_{\rm IR}$ from the 
24$\mu$m emission. Specifically, as already described in detail in 
\citet{marchesini10}, we used the infrared SEDs of star-forming galaxies from 
\citet{dale02}, which allow us to derive the IR/MIR flux ratio for different 
heating levels of the interstellar environment, parameterized by 
$dM(U) \sim U^{-\alpha} dU$, where $M(U)$ is the dust mass heated by an 
intensity $U$ of the interstellar field. The total infrared luminosity 
$L_{\rm IR,\alpha}$ was computed for each template in \citet{dale02} within the 
range $1 \leq \alpha \leq 2.5$. We adopted the log-average of the resulting 
$L_{\rm IR,\alpha=1,...,2.5}$ as the best estimate for the IR luminosity. This 
approach returns SFRs in better agreement with the SFRs estimated from SED 
modeling (\citealt{franx08}; \citealt{wuyts08}) and from dust-corrected 
H$\alpha$ line fluxes (\citealt{muzzin10}), compared to the often adopted local 
luminosity-dependent approaches, which systematically overestimate SFRs by a 
factor of 4-6 (e.g., \citealt{papovich07}; \citealt{murphy09}; 
\citealt{elbaz10}; \citealt{muzzin10}). We convert the $L_{\rm 2800}$ into 
SFR$_{\rm UV,uncorr}$ using the conversion factor 
SFR$_{\rm UV,uncorr}=3.234\times10^{-10}~L_{\rm 2800}$ from \citet{kennicutt98}, 
adapted to a \citet{kroupa01} IMF by \citet{bell05}. SFR$_{\rm UV,uncorr}$ is 
the observed SFR, and is not corrected for dust extinction. The $L_{\rm IR}$ is 
converted into a SFR$_{\rm IR}$ using 
SFR$_{\rm IR}=0.98\times10^{-10}~L_{\rm IR}$ from \citet{kennicutt98}, adapted 
to a \citet{kroupa01} IMF. The total SFR of the galaxies is then derived via 
SFR$_{\rm tot}$=SFR$_{\rm UV,uncorr}+$SFR$_{\rm IR}$.

\section{Selection of the Progenitors of Local Ultra-Massive Galaxies}\label{sec-sample}

\subsection{Fixed Cumulative Number Density}\label{fcnda}

\begin{figure*}
\epsscale{1.05}
\plotone{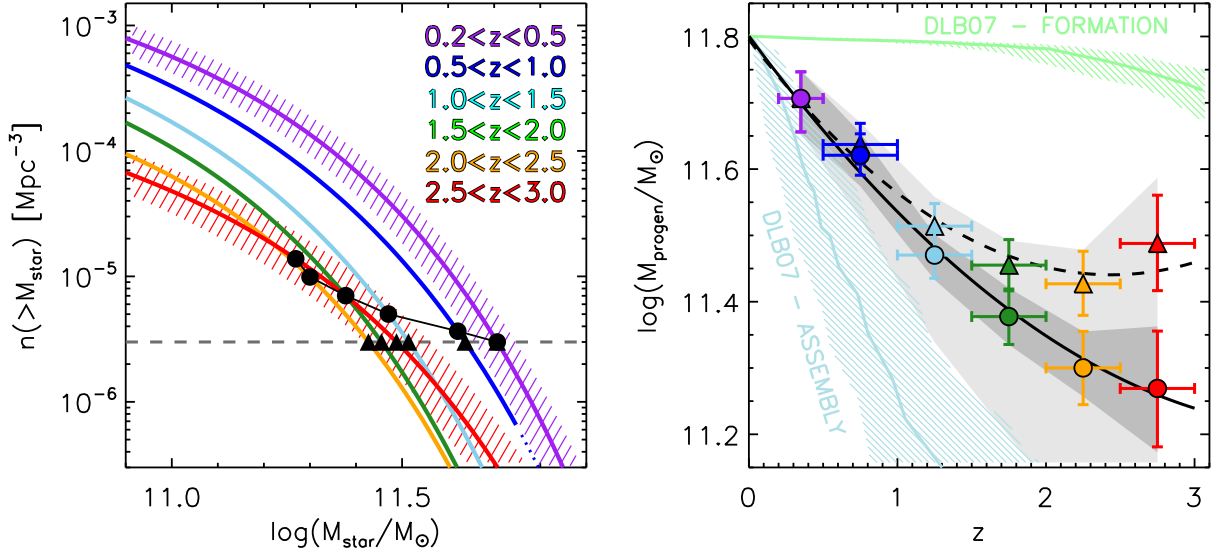}
\caption{{\bf Left panel:} Cumulative number density, $n(>M_{\rm star})$, as a 
function of stellar mass obtained by integrating the stellar mass functions 
from the UltraVISTA survey \citep{muzzin13b} at $0.2<z<3.0$. The different 
colors correspond to the different targeted redshift intervals. Filled 
triangles show the evolution in stellar mass for galaxies at fixed cumulative 
number densities of 3$\times10^{-6}$ Mpc$^{-3}$, shown by the dashed horizontal 
gray line. Filled circles show evolution in stellar mass after inclusion of the 
evolving number density for galaxy progenitors using an abundance matching 
approach (\S\ref{seaam}). The hatched regions show the 1$\sigma$ range of 
$n(>M_{\rm star})$ including Poisson errors, cosmic variance, and photometric 
redshift uncertainties. The 1$\sigma$ range of $n(>M_{\rm star})$ is plotted 
only for the lowest and highest targeted redshift intervals for clarity. 
{\bf Right panel:} Evolution of the stellar mass of the progenitors, 
$M_{\rm progen}$, as a function of redshift. Filled triangles are obtained 
using the fixed cumulative number density approach, whereas filled circles are 
derived using the semi-empirical approach using abundance matching. The error 
bars show the width of the redshift bin and the formal 1$\sigma$ uncertainty 
in stellar mass at a given cumulative number density and includes Poisson 
errors, cosmic variance, and photometric redshift uncertainties. The black 
dashed and solid curves show the evolutions of the stellar mass as a function 
of redshift (parameterized as $\log{(M_{\rm star})}=A+Bz+Cz^{2}$) derived from 
the constant cumulative number density and abundance matching approaches, 
respectively. The progenitors of $z\sim0$ UMGs (i.e., 
$\log{(M_{\rm star}/M_{\odot})}=11.8$) are selected in a narrow range of stellar 
mass around these curves. The dark gray filled region shows the 1$\sigma$ 
uncertainty on the stellar mass growth of the progenitors as selected using 
the abundance matching approach when only the errors on the observed stellar 
mass functions are considered. The light gray filled region shows instead the 
68th-percentile range in stellar mass including both the errors on the 
observed stellar mass functions and the spread from individual galaxy growth 
histories. Specifically, the 68-th percentile range was estimated by 
generating 1000 realizations of the stellar mass tracks from 1000 realizations 
of the stellar mass functions and of the galaxy accretion histories 
\citep{behroozi13b}. The color curve and hatched region represent the median 
and the 15$^{\rm th}$/85$^{\rm th}$ percentile range, respectively, of the 
assembly history (light blue) and the star-formation history (light green) of 
BCGs as simulated in \citet{delucia07}.\label{fig1}}
\end{figure*}

Following previous studies (e.g., \citealt{wake06}; \citealt{tojeiro10}; 
\citealt{brammer11}; \citealt{papovich11}; \citealt{tojeiro12}; 
\citealt{patel13}; \citealt{vandokkum13}; \citealt{leja13}; 
\citealt{muzzin13b}), we connect progenitor galaxies by requiring that they 
have the same cumulative co-moving number density of the low-redshift galaxy 
population, i.e., galaxies at $0.2<z<0.5$ with $M_{\rm star} \approx 5 \times 
10^{11}$ M$_{\odot}$. Effectively, galaxies are ranked according to their 
stellar mass and we select galaxies at different redshifts that have the same 
rank order as the low-redshift population. The implicit assumption is that rank 
order is conserved through cosmic time, or that processes that break the rank 
order do not have a strong effect on the average measured properties.

In order to apply the fixed cumulative number density selection, we derive 
cumulative number densities, $n(>M_{\rm star})$, as a function of stellar mass 
using the recently published and publicly available stellar mass functions 
from the UltraVISTA survey \citep{muzzin13b}. The contributions from Poisson 
errors, cosmic variance, and photometric redshift uncertainties to the total 
error budget of the stellar mass functions were quantified and presented in 
\citet{muzzin13b}. We refer to this paper for a detailed description of the 
derivations of the errors of the stellar mass functions. The left panel of 
Figure~\ref{fig1} shows $n(>M_{\rm star})$ in the six targeted redshift ranges 
from $z=0.2$ to $z=3.0$, namely $0.2 \le z<0.5$, $0.5 \le z<1.0$, 
$1.0 \le z<1.5$, $1.5 \le z<2.0$, $2.0 \le z<2.5$, $2.5 \le z<3.0$. Also shown 
is the total 1$\sigma$ error (including Poisson errors, cosmic variance, and 
photometric redshift uncertainties) on $n(>M_{\rm star})$ for the lowest and 
highest targeted redshift bins. Galaxies with stellar masses 
$M_{\rm star}\approx5 \times 10^{11}$~M$_{\odot}=10^{11.7}$ M$_{\odot}$ at 
$z=0.35$ have a cumulative number density of $3\times10^{-6}$ Mpc$^{-3}$ (shown 
by the horizontal dashed line in the left panel of Fig.~\ref{fig1}). We then 
trace the progenitors of these galaxies by identifying, at each redshift, the 
stellar mass for which the cumulative number density is $3\times10^{-6}$ 
Mpc$^{-3}$ (indicated in the left panel of Fig.~\ref{fig1} by the filled 
triangles). 

The stellar mass evolution for galaxies with the rank order of the low-$z$ 
population is shown in the right panel of Figure~\ref{fig1} with colored filled 
triangles. Also shown is the formal 1$\sigma$ uncertainty on the stellar mass 
of the progenitors at the given cumulative number density and including the 
contributions from Poisson errors, cosmic variance, and photometric redshift 
uncertainties. As previously done in the literature (e.g., 
\citealt{vandokkum13}), we parameterize the evolution of the stellar mass of 
the progenitors with a quadratic function of the form 
$\log{(M_{\rm progen}(z)/M_{\odot})}=A+Bz+Cz^{2}$, with $A=11.796\pm0.043$, 
$B=-0.291\pm0.076$, and $C=0.060\pm0.028$ (dashed curve in the right panel of 
Fig.~\ref{fig1}), corresponding to a value of the stellar mass in the local 
universe of $\log{(M_{\rm progen}(z=0)/M_{\odot})}=11.8$. This value is 
comparable to the typical stellar masses of local brightest cluster galaxies 
(BCGs; e.g., \citealt{lidman12}; \citealt{lin13}). 

\subsection{Semi-empirical Approach Using Abundance Matching}\label{seaam}

The fixed cumulative number density approach does not take into account 
galaxy-galaxy mergers and scatter in mass accretion histories, both altering 
the median cumulative number density of a galaxy population (\citealt{leja13}; 
\citealt{lin13}; \citealt{behroozi13a}). Therefore, to select the progenitors 
we also adopt a semi-empirical approach using abundance matching in the 
$\Lambda$CDM paradigm, accounting for mergers and scatter in mass accretion 
histories. A detailed description of this technique is presented in 
\citet{behroozi13b} (and references therein), who also made publicly available 
the code implementing this 
technique.\footnote{http://code.google.com/p/nd-redshift/} Briefly, the galaxy 
cumulative number density at redshift $z_{\rm 1}$ is converted to a halo mass 
with equal cumulative number density using peak halo mass functions. Then, for 
halos at that mass at $z_{\rm 1}$, the masses of the most-massive progenitor 
halos at $z_{\rm 2}>z_{\rm 1}$ are recorded using to the halos' mass accretion 
histories. Finally, the median halo progenitor mass at $z_{\rm 2}$ is converted 
back into cumulative number densities using the halo mass function at 
$z_{\rm 2}$. 

Similarly to the fixed cumulative number density approach, the progenitors of 
the low-$z$ population of very massive galaxies are traced by identifying, at 
each redshift, the stellar mass for which the evolving cumulative number 
density (determined with the semi-empirical approach using abundance matching) 
intersects the cumulative number density curves derived from the UltraVISTA 
stellar mass functions (plotted as filled circles in the left panel 
Fig.~\ref{fig1}).

The resulting stellar mass evolution with the corresponding 1$\sigma$ 
uncertainty is shown in the right panel of Figure~\ref{fig1} with colored 
filled circles. Errors on the stellar mass of the progenitors were derived in 
the same way as described in \S\ref{fcnda}. Adopting the same parameterization 
as for the fixed cumulative number density approach, we find 
$A=11.801\pm0.038$, $B=-0.304\pm0.054$, and $C=0.039\pm0.014$, also implying a 
value of the stellar mass in the local universe of 
$\log{(M_{\rm progen}(z=0)/M_{\odot})}=11.8$ (continuous curve in the right 
panel of Fig.~\ref{fig1}).

\section{Evolution of the Progenitors' Properties}\label{sec-results}

Here we study and present the evolution as a function of cosmic time of the 
properties of the progenitors of local ultra-massive 
($\log{(M_{\rm star}/M_{\odot})}=11.8$) galaxies (UMGs, hereafter) since $z=3$. 
For a given redshift, we select the progenitors of $z\sim0$ UMGs within a bin 
of size $\sim0.2$~dex in stellar mass centered on $M_{\rm progen}(z)$, following 
the approach adopted by \citet{patel13}. The actual boundaries of the bins are 
adjusted such that the median mass is close to the value given by 
$M_{\rm progen}(z)$. Given the steepness of the stellar mass function, in 
practice this results in selecting galaxies at 
$(\log{M_{\rm progen}(z)/M_{\odot}})^{+0.15}_{-0.07}$. 

\subsection{Stellar Mass Evolution}

As shown by the dashed curve in the right panel of Figure~\ref{fig1}, the 
evolution in stellar mass of the progenitors of $z\sim0$ UMGs is quite small 
when the progenitors are selected using the fixed cumulative number density 
method, i.e., $0.33\pm0.15$~dex (i.e., a factor of $\sim$2.1$^{+1.0}_{-0.6}$) 
from $z=3.0$ to $z=0$. Therefore, about half of the stellar mass in local UMGs 
was assembled at $z>3$, i.e., in the first 2~Gyr of cosmic history. If the 
progenitors are instead selected with the semi-empirical approach using 
abundance matching, the evolution in stellar mass is much more significant, 
$0.56\pm0.21$~dex, $0.45\pm0.13$~dex, and $0.27\pm0.08$~dex from $z=3.0$, 
$z=2.0$, and $z=1.0$, respectively, to $z=0$. Therefore, the progenitors have 
grown by a factor of $\sim$3.6$^{+2.3}_{-1.4}$ over the last 11.4~Gyr from 
$z=3.0$ to $z=0$, with only about a fourth of the stellar mass in local UMGs 
having been assembled by $z\sim3$. If the 68th-percentile range in stellar 
mass (shown in Figure~\ref{fig1} as a light gray filled region) resulting from 
the scatter in the progenitors' number density is included in the error 
budget, the uncertainty on the inferred growth would be larger by a factor up 
to $\sim$1.7, i.e., $0.56^{+0.35}_{-0.25}$~dex, $0.45^{+0.16}_{-0.20}$ dex, and 
$0.27^{+0.08}_{-0.12}$~dex from $z=3.0$, $z=2.0$, and $z=1.0$, respectively, to 
$z=0$.

The right panel of Figure~\ref{fig1} also shows the predicted median and 
15$^{\rm th}$/85$^{\rm th}$ percentiles of the `formation' (light blue) and 
`assembly' (light green) histories of BCGs from the theoretical study using 
semi-analytic techniques presented in \citet{delucia07}. The formation history 
of BCGs is defined as the sum of the 
stellar masses in all progenitors at each time, whereas their assembly history 
corresponds to the stellar mass of the main branch. As highlighted in 
\citet{delucia07}, there is a very small scatter in the formation histories of 
BCGs, with most of their stars having formed at $z>$3 ($\sim$90\% of the stars 
formed at $z>2.5$). The assembly histories exhibit a much larger scatter with 
a much later assembly time. Specifically, the fraction of mass in the main 
progenitor varies between 15\% and 40\% at $z=1$, and between 40\% and 70\% at 
$z=0.5$ \citep{delucia07}. Figure~\ref{fig1} shows that the inferred 
evolutions in stellar mass of the progenitors of $z\sim0$ UMGs (solid or 
dashed curves) are faster than the model-predicted assembly histories, but 
much slower than the predicted formation histories of BCGs. In other words, in 
semi-analytic models most of the stars were not formed in the main branch, but 
star formation happened very early in low- to intermediate-mass halos (with 
stellar masses $\sim$a few $\times$10$^{10}$~M$_{\odot}$), which were accreted 
steadily over time following the assembly history of the dark matter. We note 
that, as the median cumulative number density of the progenitors increases 
with redshift as $\sim(0.2 \Delta z)$ \citep{behroozi13b}, using the fixed 
cumulative number density approach would preferentially select progenitors 
which potentially had an unusually rapid assembly at earlier times. This 
should be kept in mind in the comparison between the model-predicted assembly 
history of BCGs with the inferred growth in stellar mass of the progenitors of 
local UMGs selected with the fixed cumulative number density approach. At 
$z=3$, the progenitors of UMGs have a typical observed stellar mass of 
$\log{(M_{\rm star}/M_{\odot})}\approx11.2$, a factor of $\sim5$ larger than the 
model-predicted stellar mass assembled by $z=3$, implying a predicted 
abundance of BCGs' progenitors at $z=3$ significantly larger than observed. 
We note that the dominant source of uncertainty in the stellar mass evolution 
of the progenitors is due to the scatter in the progenitors' cumulative number 
density. When this source of error is included, as shown by the light gray 
filled region in the right panel of Figure~\ref{fig1}, the disagreement 
between the observed stellar growth of local UMGs and the model-predicted 
assembly history of BCGs becomes less significant, and the two marginally 
agree at the $\sim$2$\sigma$ level.

The disagreement between observed and model-predicted evolutions of the 
stellar mass content of the progenitors of local UMGs is another manifestation 
of the previously highlighted tensions between observed and model-predicted 
stellar mass functions. Specifically, semi-analytic models were found to 
significantly overpredict the observed number density of galaxies below 
$10^{11}$~M$_{\odot}$ at $z\gtrsim1$-1.5 and to underpredict the number density 
of massive galaxies at $z>2$ (e.g., \citealt{marchesini09}), with low-mass 
galaxies in the stellar mass range $10^{9}-10^{11}$~M$_{\odot}$ forming too 
early in the models and being too passive at later times (e.g., 
\citealt{fontanot09}; see also \citealt{henriques12}). We note that, whereas 
the typical errors on stellar masses are not large enough to explain the 
disagreement at the high-mass end between the observed and the model-predicted 
stellar mass functions at $z \gtrsim 1$ (e.g., \citealt{ilbert13}), the 
disagreement between the inferred evolution in stellar mass of the progenitors 
of local UMGs and the model-predicted assembly history of BCGs shown in 
Figure~\ref{fig1} could potentially be further reduced (although at most only 
partly) if the model predictions were to be convolved by the typical 
statistical errors in the stellar masses of UltraVISTA.

Whereas the inferred growth in stellar mass from $z=3$ is quite different 
depending on the adopted method to select the progenitors, the derived 
evolution in the stellar population parameters (e.g., rest-frame colors, level 
of star formation activity, stellar ages, extinction, rest-frame optical 
absolute magnitudes, and mass-to-light ratios) is qualitatively similar. 
Moreover, observationally, recent works have robustly revealed that most 
early-type galaxies show tidally disrupted features indicative of their past 
merging histories (e.g., \citealt{malin83}; \citealt{vandokkum05}; 
\citealt{tal09}; \citealt{janowiecki10}; \citealt{sheen12}), with evidence of 
major dry (i.e., dissipationless) merging being an increasingly important 
mechanism for the evolution of galaxies with $M_{\rm star}>10^{11.4}$~M$_{\odot}$ 
(e.g., \citealt{kim13}). Contrary to the fixed cumulative number density 
method, the semi-empirical approach using abundance matching accounts for 
mergers, making the latter a more desirable approach to constrain the evolution 
of the median cumulative number density for a galaxy population 
\citep{behroozi13b}. Therefore, in the sections below, we present only the 
evolution of the stellar population properties of the progenitors of $z\sim0$ 
UMGs when selected with the semi-empirical approach using abundance matching. 
For completeness, Appendix~\ref{appendixA} presents the evolution of the 
stellar population properties of the progenitors of local UMGs when selected 
with the fixed cumulative number density method.

\subsection{Evolution in Rest-frame $U-V$ and $V-J$ Colors}

\begin{figure*}
\epsscale{1.1}
\plotone{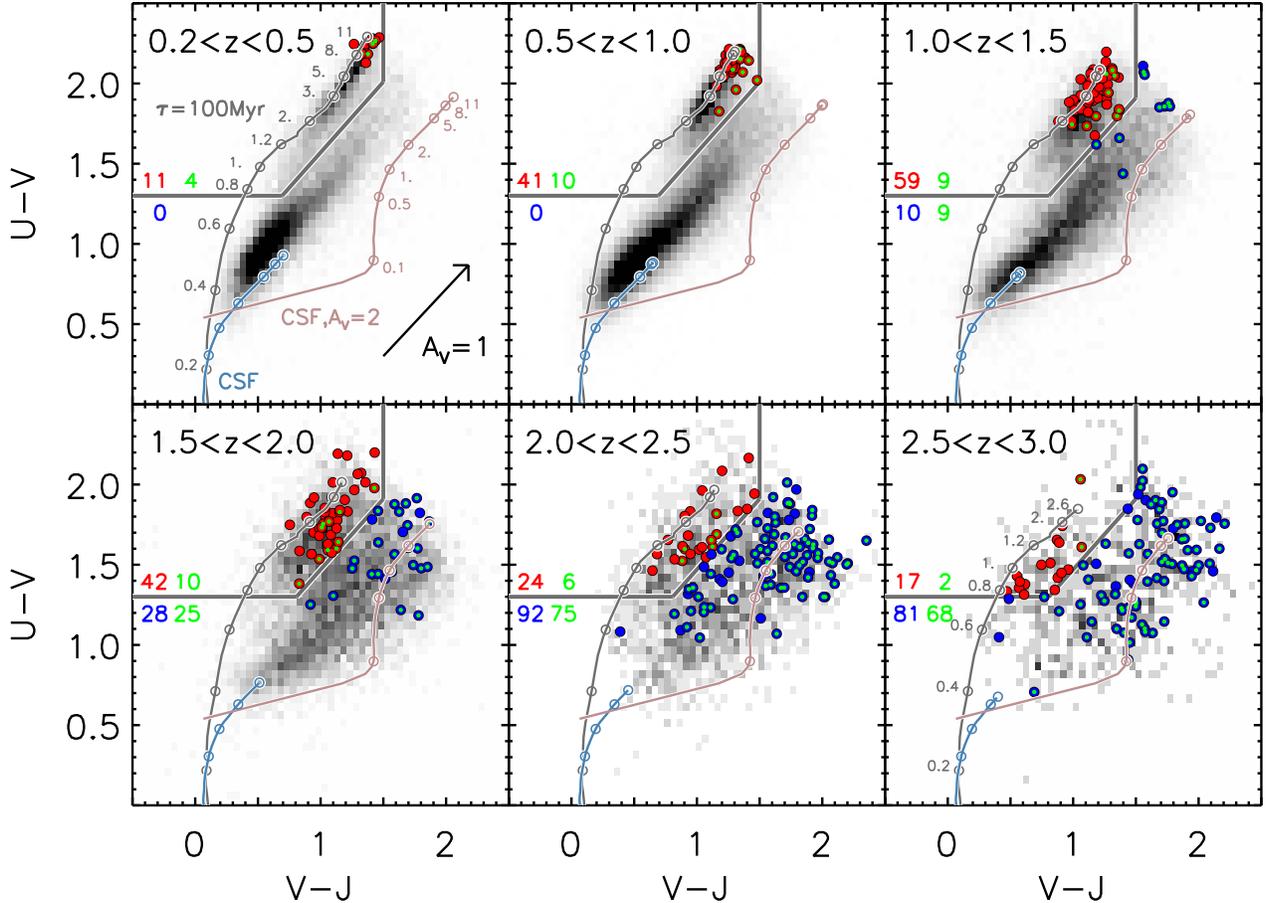}
\caption{Rest-frame $U-V$ versus $V-J$ diagram in the targeted redshift 
intervals between $z=0.2$ and $z=3.0$ for the progenitors of local UMGs 
selected with the semi-empirical approach using abundance matching. The cuts 
used to separate star forming from quiescent galaxies from \citet{muzzin13b} 
are shown as the solid dark gray lines. Red and blue filled circles highlight 
the quiescent and star-forming galaxies among the progenitors' sample. Listed 
are the numbers of quiescent (red) and star-forming (blue) progenitors in each 
redshift interval. The grayscale representation of the overall galaxy 
population above the 95\% mass-completeness limit from the UltraVISTA survey 
is also plotted in each redshift interval. Small green circles show the 
progenitors detected in the {\it Spitzer}-MIPS 24$\mu$m data with 
signal-to-noise $S/N>5$. The numbers of MIPS-detected galaxies among the 
quiescent and star-forming progenitors are listed in green. Most of the 
star-forming galaxies are robustly detected in MIPS, whereas only a small 
fraction of quiescent galaxies is detected at 24$\mu$m. Color evolution tracks 
of \citet{bruzual03} models are also shown: an exponentially declining SFH 
with no dust ($\tau$=100~Myr; dark gray), a constant SFH with no dust (CSF; 
light blue), and the same CSF model with $A_{\rm V}=2$~mag of extinction 
(light brown). The evolution tracks are plotted up to the maximum allowed age 
of the universe corresponding to the lower limit of each redshift interval. 
The empty circles represent the model colors at the specified ages (in Gyr). 
The dust vector indicates an extinction of $A_{\rm V}=1$~mag for a 
\citet{calzetti00} extinction curve. The population of progenitors of local 
UMGs shows a range in properties at $2.5<z<3.0$, typical of a heterogeneous 
population mostly dominated by dusty ($A_{\rm V}=1.5-2$~mag) star-forming 
galaxies, but also including post-starburst quiescent galaxies. By $z=0.35$, 
such heterogeneous population has turned into a homogeneous population of 
maximally old, quiescent galaxies with very similar colors. 
\label{fig2}}
\end{figure*}

The rest-frame $U-V$ versus $V-J$ color-color diagram (hereafter the $UVJ$ 
diagram) has become, in recent years, a common tool to separate quiescent from 
star-forming galaxies for its ability to separate red galaxies that are 
quiescent from reddened (i.e., dust obscured) star-forming galaxies (see, 
e.g., \citealt{labbe06}; \citealt{wuyts07}; \citealt{williams09}; 
\citealt{brammer11}; \citealt{patel11}; \citealt{whitaker11}; 
\citealt{muzzin13b}). Figure~\ref{fig2} shows the evolution of the rest-frame 
$U-V$ and $V-J$ colors of the progenitors of UMGs since $z=3$ (colored filled 
circles), along with the evolution of the rest-frame $U-V$ and $V-J$ colors of 
the overall galaxy population (grayscale representation) above the 95\% 
mass-completeness limit from the UltraVISTA survey. To distinguish between 
star-forming and quiescent galaxies we use the same box adopted in 
\citet{muzzin13b} for the study of the evolution of the stellar mass functions 
of star-forming and quiescent galaxies since $z=4$. The adopted box is plotted 
in Figure~\ref{fig2}, with the quiescent and star-forming progenitors 
highlighted in red and blue, respectively. Color evolution tracks from the 
stellar population synthesis models of \citet{bruzual03} are also plotted in 
Figure~\ref{fig2} for different assumed star-formation histories (SFHs) and 
amount of dust extinction: an exponentially declining SFH with an e-folding 
timescale of $\tau$=100~Myr and no dust extinction ($\tau_{\rm 100}$; dark gray 
track), a constant SFH with no dust (CSF; light blue), and the same CSF model 
with $A_{\rm V}=2$~mag of extinction (light brown). The color tracks are 
plotted up to the maximum allowed age of the universe corresponding to the 
lower limit of each redshift interval. We note that the $\tau_{\rm 100}$ color 
track is very similar to the color track of a single stellar population, with 
the latter reaching the same $U-V$ and $V-J$ colors of the $\tau_{\rm 100}$ 
model at slightly younger (by $\sim0.2$~Gyr) ages. 

At $0.2<z<0.5$, the progenitors of UMGs constitute a homogeneous population 
with similar rest-frame $U-V$ and $V-J$ colors, typical of 
quiescent galaxies with old (age$>8$~Gyr) stellar populations (from the 
comparison with the $\tau_{\rm 100}$ color track). As the progenitors of UMGs 
are followed to higher redshifts, such population becomes increasingly 
diversified. Out to $z=1$, the progenitors are still all quiescent galaxies 
and constitute a homogeneous population, while at $z>1$ the contribution 
from star-forming galaxies progressively increases, with the population of the 
progenitors being dominated by star-forming galaxies at $2<z<3$. 
Quantitatively, the fraction of quiescent galaxies in the progenitors of local 
UMGs is seen to decrease from 100\% at $z \le 1$, to $\sim$86\% at $z=1.25$, 
and down to $\sim$17\% by $z=2.75$. 

The location of the star-forming progenitors in the $UVJ$ diagram clearly 
indicates that these galaxies are quite dusty, with typical extinction in the 
visual band of $A_{\rm V}\sim1.5-2$~mag (inferred from the comparison with the 
color track of a CSF model with $A_{\rm V}=2$~mag). The dusty nature of 
the star-forming progenitors can be further tested with {\it Spitzer}-MIPS 
24$\mu$m data. For the targeted redshift range, the MIPS~24$\mu$m probes the 
rest-frame mid-infrared (MIR; $\sim$6 and 20$\mu$m at $z=3$ and $z=0.2$, 
respectively), offering a powerful method of separating two basic SED types: 
evolved quiescent systems (whose stellar emission drops drastically at 
$\lambda_{\rm rest} >$2$\mu$m), and active dusty galaxies, i.e., those that are 
powered by either star formation or active galactic nuclei (AGNs), both of 
which produce substantial MIR emission (e.g., \citealt{webb06}). Specifically, 
in pure starburst galaxies, the MIR emission is dominated by features from 
polycyclic aromatic hydrocarbons (PAH) that are strong relative to the 
underlying dust continuum, which only begins to rise above $\sim10$$\mu$m 
\citep{draine07}. The hard radiation field of an AGN destroys PAH carriers, 
and the continuum emission from hot small dust grains is strong throughout the 
MIR \citep{genzel00}. 

The progenitors of UMGs detected with a signal-to-noise ratio $S/N>5$ in the 
MIPS 24$\mu$m data are shown in Figure~\ref{fig2} by green filled circles. On 
average, $\sim$84\% of the star-forming progenitors are robustly detected with 
$S/N>5$ in MIPS, with a fraction of 24$\mu$m-detected sources as high as 90\% 
at $z<2$, supporting the dusty nature of the star-forming progenitors. We note 
that the adopted MIPS flux limit (35$\mu$Jy, 5$\sigma$) translates into a 
large range of MIR luminosities sampled at a broad range of rest-frame 
wavelengths over the large considered redshift range, preventing the detection 
rate alone to be an ideal proxy of the level of obscured star-formation 
activity as a function of redshift, since larger SFRs are needed in order to 
be detected at higher redshifts. Therefore, the roughly constant detection 
rate of the star-forming progenitors over the targeted redshift interval is a 
strong evidence for increasing levels of the dust-enshrouded activity with 
redshift in the star-forming progenitors of UMGs (see a more quantitative 
analysis of the evolution of the SFR and specific-SFR in \S\ref{ssfr}). 

\begin{figure*}
\epsscale{1.1}
\plotone{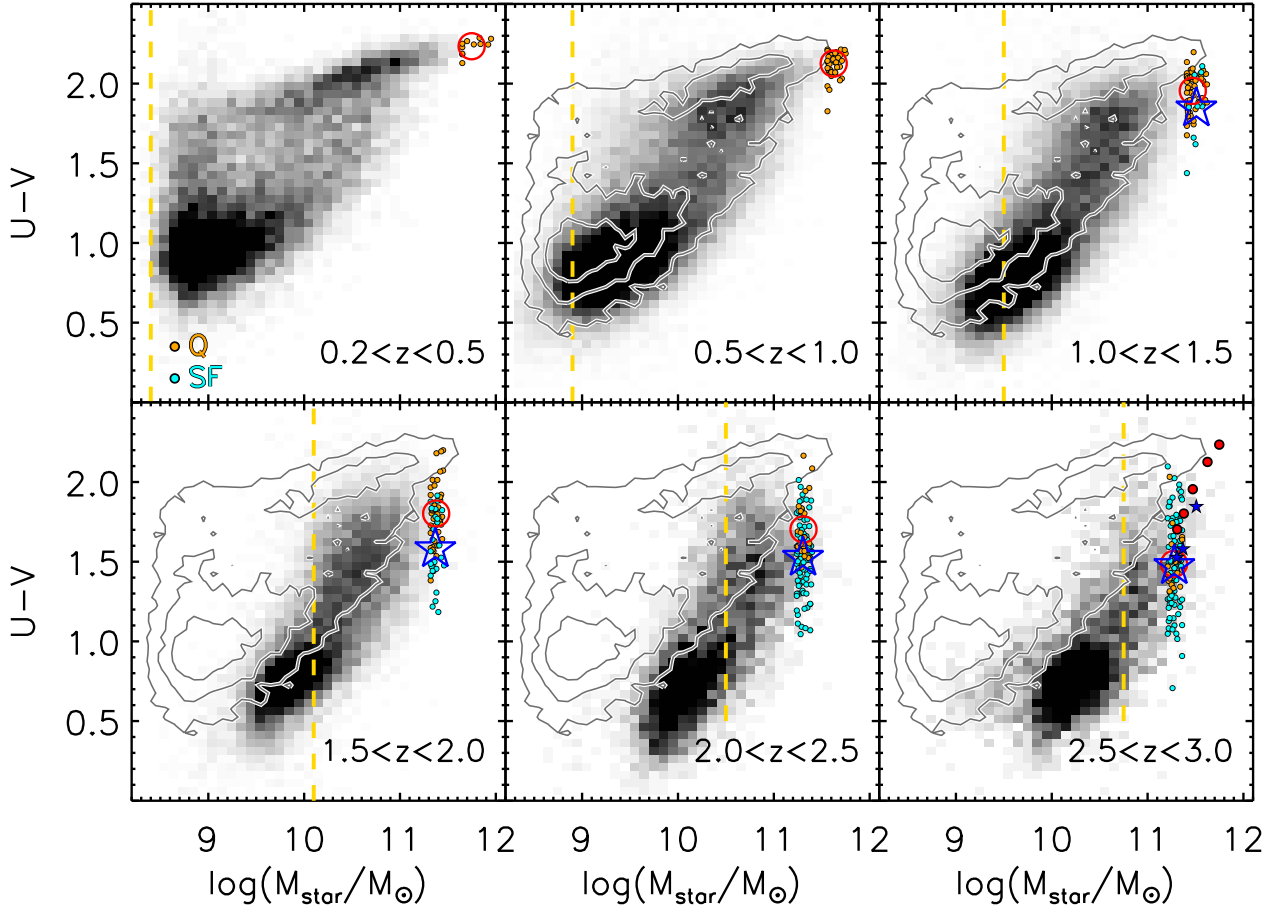}
\caption{Rest-frame $U-V$ color versus stellar mass diagram. The grayscale 
representation shows the $U-V$ versus $M_{\rm star}$ diagram for the overall 
galaxy population from the UltraVISTA survey in each redshift interval, with 
the vertical dashed yellow line representing the 95\% completeness in stellar 
mass. The $U-V$ versus $M_{\rm star}$ diagram at $0.2 \leq z<0.5$ is overplotted 
in all panels at $z>0.5$ as gray contours to highlight the evolution as a 
function of redshift of the overall galaxy population in the $U-V$ versus 
$M_{\rm star}$ diagram. Orange and cyan small filled circles represent the 
quiescent and star-forming progenitors of UMGs. The open red circle and blue 
star represent the median value of the quiescent and star-forming progenitors, 
respectively. In the panel corresponding to the $2.5 \le z < 3.0$ redshift 
interval, the small filled red circles and blue stars represent the lower 
redshift median values of the quiescent and star-forming progenitors, 
respectively, to highlight their evolution with redshift in the color-stellar 
mass diagram. The star-forming progenitors have never lived in the blue cloud 
since $z=3$, and they are as red as the quiescent red-sequence galaxies at all 
redshifts.\label{fig3}}
\end{figure*}

In contrast to the star-forming progenitors, only $\sim$21\% of the quiescent 
progenitors are on average detected at 24$\mu$m with $S/N>5$, with this 
fraction ranging from $\sim$36\% at $z=0.35$ to $\sim$12\% at $z=2.75$. This 
detection rate is quantitatively consistent with the MIPS detection rate found 
in a sample of quiescent galaxies at $0.3<z<2.5$ with 
$\log{(M_{\rm star})/M_{\odot}}>10.3$ selected from the 3D-HST survey 
\citep{fumagalli13}. While the MIPS detection rate in the quiescent 
progenitors is a factor of $\sim$4 smaller than in the star-forming 
progenitors, its not-zero value is potentially indicative of low levels of 
obscured star formation or AGN contamination. We see from Figure~\ref{fig2} 
that the MIPS detection in the quiescent progenitors have preferentially 
redder $V-J$ colors, lying closer to the line separating quiescent from 
dusty star-forming galaxies, suggesting the possibility of these galaxies 
being in their terminal phases of star formation and on their ways to become 
fully quenched systems. If these systems are indeed transitioning systems in 
their final approaches to the quiescent population, it is tempting to 
interpret their MIPS detections as potentially due to obscured AGNs 
responsible for the quenching of their star-formation activities, given the 
increasing fraction of AGN activity with increasing stellar mass found both in 
the local universe (e.g., \citealt{kauffmann03b}) and out to $z\sim2.5$ (e.g., 
\citealt{papovich06}; \citealp{kriek07}). We note however that the 
investigation of the origin of the MIPS fluxes in the sample of quiescent, 
lower-mass galaxies in \citet{fumagalli13} appears to indicate that the low 
MIPS fluxes can be fully accounted by processes unrelated to obscured AGNs and 
on-going star formation, such as cirrus dust heated by old stellar populations 
and circumstellar dust embedding asymptotic giant branch stars. Related to the 
dust content, we finally note that the comparison in the $UVJ$ diagram of the 
quiescent progenitors and the evolution tracks indicates an increasing amount 
of dust extinction with increasing redshift also in the quiescent progenitors 
of UMGs, especially at $z>1-1.5$ (see also \S\ref{sec-dust}). 

\subsection{Evolution in the $U-V$ versus $M_{\rm star}$ diagram}\label{cmd}

As highlighted in the previous section, Figure~\ref{fig2} shows that the 
progenitors of UMGs at $z<1$ constitute a homogeneous population, with 
very similar rest-frame $U-V$ and $V-J$ colors, typical of quiescent and old 
stellar populations. Two additional remarkable results are evident from 
Figure~\ref{fig2}. First, the scatter in the rest-frame $U-V$ color of the 
quiescent progenitors becomes progressively larger with increasing redshift, 
especially at $z>1.5$. Second, the star-forming progenitors show similar 
rest-frame $U-V$ colors compared to the quiescent progenitors, indicative that 
the star-forming progenitors never lived on the blue cloud since $z=3$. Even 
more remarkable is that a significant fraction of star-forming progenitors in 
the highest redshift bin show rest-frame $U-V$ colors redder than the 
quiescent progenitors, in addition to significantly redder rest-frame $V-J$ 
colors, suggesting that the star-forming progenitors at $z\sim3$ of local UMGs 
may be actually {\it redder} than the quiescent progenitors. 

These results can be better seen in Figure~\ref{fig3}, which shows the 
rest-frame $U-V$ versus $M_{\rm star}$ diagram (or color-mass diagram; CMD, 
hereafter) of the quiescent and star-forming progenitors of UMGs. 
Figure~\ref{fig3} also shows the evolution in the CMD of the overall galaxy 
population (grayscale representation) from the UltraVISTA survey. 

At $0.2 \le z<0.5$, the progenitors of local UMGs are the reddest objects and 
have similar rest-frame $U-V$ colors, with a very small intrinsic scatter 
($\sim0.04$~mag). To measure the intrinsic scatter ($\sigma_{\rm int}$), we 
corrected the observed scatter ($\sigma_{\rm obs}$) for the scatter introduced 
by photometric errors ($\sigma_{\rm phot}$) using the formula 
$(\sigma_{\rm obs})^{2}=(\sigma_{\rm int})^{2}+(\sigma_{\rm phot})^{2}$. Following 
\citet{whitaker10}, for each progenitor, the observed SED fluxes were 
perturbed by a normally distributed, pseudo-random number from a Gaussian 
distribution with a mean of zero and a standard deviation of the photometric 
error for each respective filter. From these perturbed flux values, we 
generated 100 simulated catalogs, and re-determined $U-V$, refitting the 
photometric redshifts. The uncertainty in $U-V$ color for each progenitor is 
estimated from the biweight sigma of the its $U-V$ distribution. At each 
redshift interval, we estimated the values of $\sigma_{\rm phot}$ for each 
sample of quiescent and star-forming progenitors by taking the average of the 
biweight sigma for the progenitors in the considered sample. The average 
uncertainty in $U-V$ colors for the sample of quiescent progenitors ranges 
from 0.02~mag at $z=0.35$ to 0.06~mag at $z=2.75$, while for the sample of 
star-forming progenitors it ranges from 0.03~mag at $z=1.25$ to 0.12~mag at 
$z=2.75$.

The derived scatter of the progenitors at $z=0.35$ is quantitative consistent 
with the tightness of the red sequence in the local universe 
($\sim$0.03$\pm$0.02 mag; e.g., \citealt{bower92}). While at 
$0.5 \le z < 1.0$ the intrinsic scatter in the $U-V$ color is still quite small 
($\sim0.07$~mag), consistent with the scatter of $\sim$0.08$\pm$0.03 mag from 
several studies done in galaxy clusters out to $z\sim0.8$ 
(\citealt{vandokkum00}; \citealt{holden04}; \citealt{mcintosh05}; 
\citealt{mei09}; \citealt{ruhland09}; \citealt{vandokkum08}), the scatter in 
the rest-frame $U-V$ color of the quiescent progenitors becomes progressively 
larger with increasing redshift, i.e., $\sim0.11$~mag at $1.0 \le z < 1.5$ and 
$\sim0.17$~mag at $z>1.5$. The scatter and its evolution is in good 
agreement with the evolution of the rest-frame $U-V$ scatter derived from the 
NEWFIRM Medium-Band Survey (NMBS) in a mass-complete sample of quiescent 
galaxies at $0.2<z<2.2$ with $\log{(M_{\rm star}/M_{\odot})}>11$ 
\citep{whitaker10}. Our work shows, for the first time, the increasing scatter 
with redshift of the rest-frame $U-V$ color for a sample of quiescent 
progenitors of local UMGs rather than for samples selected above a fixed 
stellar mass at all redshifts and which are likely not evolutionarily 
connected. 

Figure~\ref{fig3} clearly demonstrates that the star-forming progenitors have 
never lived in the blue star-forming cloud in the last 11.4~Gyr, i.e., since 
$z=3$. Indeed, the star-forming progenitors appear to be as red as the 
quiescent galaxies at all redshifts. In the highest redshift bin, i.e., at 
$z=2.75$, quiescent and star-forming progenitors have similar median 
$U-V$ colors, although there is a significant fraction of star-forming 
progenitors with redder colors compared to most of the quiescent progenitors. 
The intrinsic scatter in $U-V$ color of the star-forming progenitors is much 
larger than the quiescent progenitors, i.e, $\sim0.28$~mag at $z=2.75$ and 
$\sim0.22$~mag at $z<2.5$. As shown in the bottom-right panel of 
Figure~\ref{fig3}, the median $U-V$ color of the star-forming progenitors 
does not seem to evolve much in the $\sim1.3$~Gyr from $z=2.75$ to $z=1.75$, 
whereas the median $U-V$ color of the population of quiescent progenitors 
becomes redder as it matures from a young, post-starburst to a more evolved 
population (see also \S\ref{ages} and \ref{seds}). 

The large scatter in $U-V$ of the quiescent progenitors at $z>1.5$ is evidence 
that the massive end of the local red sequence is in the process of assembly 
between $z=3$ and $z=1.5$. At $z=2.75$, $\sim$15\% of the progenitors of UMGs 
have already quenched, most of which only recently (from Fig.~\ref{fig2}; see 
also \S\ref{ages}, \ref{ssfr}, and \ref{seds}). The majority of the 
progenitors at $z=2.75$ are star bursting. Then, the quiescent progenitors age 
with cosmic time and redden, while most of the star-forming progenitors quench 
in the $\sim2.6$~Gyr from $z=2.75$ to $z=1.25$. The few remaining star-forming 
progenitors at $z=1.25$ complete their migration to the quiescent population 
before $z=1$, i.e., in less than 1~Gyr. Therefore, the massive end of the 
quiescent red sequence is fully formed by $z=1$, and it ages thereafter in the 
following $\sim7.7$~Gyr from $z=1$ to $z\sim0$.

\subsection{Evolution in Stellar Population Properties}\label{sec-prop}

We now focus on the evolution with redshift of the stellar population 
properties of the progenitors of UMGs as derived from the modeling of the 
UV-to-8$\mu$m SEDs. 

\subsubsection{Stellar ages}\label{ages}

\begin{figure}
\epsscale{1.1}
\plotone{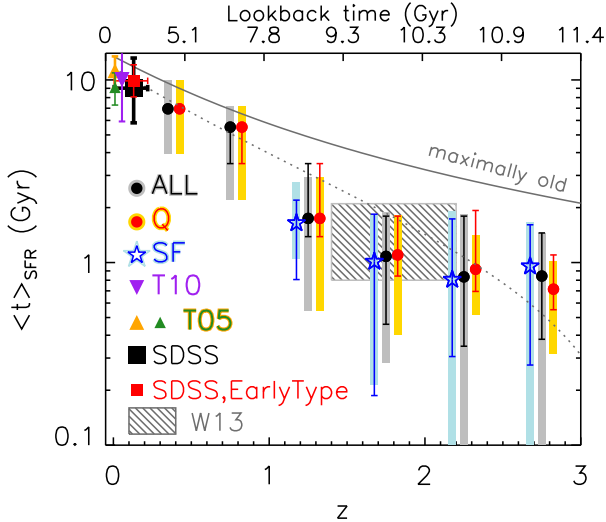}
\caption{Evolution with redshift of the stellar age 
$\langle t \rangle_{\rm SFR}$ derived from the modeling of the UV-to-8$\mu$m 
SEDs of the progenitors of local UMGs selected with the semi-empirical 
approach using abundance matching (black filled circles). Red filled circles 
and blue empty stars represent the quiescent and star-forming galaxies among 
the progenitors' sample, respectively. The error bars show the upper and lower 
ranges including 68\% of the populations, while the gray, orange, and light 
blue filled regions represent the mean 1$\sigma$ errors in the estimated 
stellar ages for the whole progenitors' sample, the quiescent, and 
star-forming progenitors, respectively. The continuous curve represents the 
age of the universe at the corresponding redshift, while the dotted curve is 
the continuous curve shifted down by 1.8~Gyr, and going through the 
measurements at $z\sim0$. The red and blue symbols are slightly offset along 
the x-axis (in opposite directions) for clarity. Also plotted are the average 
age of quiescent galaxies at $1.4<z<2.2$ with 
$\log{(M_{\rm star}/M_{\odot})}>10.5$ selected from the 3D-HST survey (hatched 
gray box; \citealt{whitaker13}); the ages of early-type galaxies at $z\sim 0$ 
with $\log{(M_{\rm star}/M_{\odot})}\approx11.8$ in high- and low-density 
environments (orange and green filled upward triangles; \citealt{thomas05}); 
the age of early-type galaxies at $0.05<z<0.06$ with 
$\log{(M_{\rm dyn}/M_{\odot})}\approx11.8$ from the SDSS (purple filled downward 
triangle; \citealt{thomas10}); the typical age of all galaxies at $z\sim 0.13$ 
with $\log{(M_{\rm star}/M_{\odot})}\approx11.8$ from the SDSS (black filled 
square; \citealt{gallazzi05}); and the typical age of early-type galaxies at 
$z\sim 0.13$ with $\log{(M_{\rm star}/M_{\odot})}\approx11.8$ from the SDSS 
(filled red square; \citealt{gallazzi06}). Whereas the median age of the 
star-forming progenitors remains roughly constant from $z=3$ to $z=1$ with 
ages$\sim$0.8-1.6~Gyr (although with large uncertainties), the median stellar 
age of the quiescent progenitors increases steadily from 
$\langle t \rangle_{\rm SFR}\sim0.7$~Gyr at $z=2.75$ to 
$\langle t \rangle_{\rm SFR}\sim7$~Gyr at $z=0.35$, reaching an age of 
$\sim$9-10~Gyr by $z\sim0$.\label{fig4}}
\end{figure}

Figure~\ref{fig4} shows the evolution with redshift of the stellar age derived 
from SED modeling of the progenitors of UMGs. As stellar age of the galaxy, we 
adopt the SFR-weighted mean age of the stellar population, 
$\langle t \rangle_{\rm SFR}$, as defined in \citet{natascha04}. When modeling 
SEDs, the best-fit age is strictly speaking the time elapsed since the onset 
of star formation. However, in exponentially declining SFHs, galaxies are 
continually forming stars, and therefore $\langle t \rangle_{\rm SFR}$ 
corresponds more closely to the age of the stars contributing the bulk of the 
stellar mass (and dominating the light of the galaxy). 

The typical age of the star-forming progenitors is roughly constant at 
$\langle t \rangle_{\rm SFR} \approx$0.8-1~Gyr from $z=3$ to $z=1.5$, although 
with a large range of ages spanned at each redshift, and it increases to 
$\sim$1.6~Gyr by $z=1.25$. The 68\% of the population of star-forming 
progenitors at $z>1.5$ is bracketed by a lower limit in age of 
$\sim$0.2-0.3~Gyr and an upper limit in age of $\sim$1.6-1.8~Gyr. On the 
contrary, the stellar age of the quiescent progenitors increases steadily from 
$z=2.75$ (with a median age of $\sim$0.7 Gyr) to $z=0.35$ (with a median age of 
$\sim$7~Gyr). The quiescent progenitors also show a smaller spread in stellar 
age at each redshift compared to the star-forming progenitors. Specifically, 
at $z>1$, the 68\% of the quiescent population is bracketed by a lower limit 
in age $\sim$20\% smaller than the median value (compared to 60\%-80\% for 
the star-forming progenitors). We note that the stellar ages derived for the 
star-forming progenitors at $z>2$ are significantly more uncertain compared to 
the estimated ages of the quiescent progenitors. As indicated by the orange and 
light blue filled regions in Figure~\ref{fig4}, the formal error on the 
stellar ages of the star-forming progenitors is 0.7-1.1~Gyr at $z>2$, a factor 
of $\sim2$-2.5 larger than for quiescent progenitors at the same redshifts. 

Figure~\ref{fig4} clearly shows the aging of the quiescent progenitors of 
local UMGs with decreasing redshift. This result is also evident from the 
$UVJ$ diagram shown in Figure~\ref{fig2}. Specifically, at $2.5<z<3.0$, the 
quiescent progenitors mostly populate the lower-left corner of the quiescent 
box. From the comparison with the $\tau_{\rm 100}$ color evolution track (dark 
gray curve in Fig.~\ref{fig2}), this region in the quiescent box of the $UVJ$ 
diagram corresponds to ages of 0.8-1.2~Gyr, and even younger ages for 
non-negligible amounts of dust extinction (see \S\ref{sec-dust}), in good 
agreement with the ages derived from SED modeling. With decreasing redshift, 
we see from Figure~\ref{fig2} that the population of quiescent progenitors of 
UMGs progressively migrates toward the upper-right corner of the quiescent 
box, clearly showing the aging of the stellar population of the quiescent 
progenitors as a function of cosmic time. 

Whereas the above results are photometrically derived, they agree well with 
recent results showing the spread in spectroscopically-derived ages of 
quiescent galaxies at $1.4<z<2.2$ from the 3D-HST dataset \citep{whitaker13}. 
Specifically, in \citet{whitaker13}, massive 
($\log{(M_{\rm star}/M_{\odot})}>10.5$), quiescent galaxies at $1.4<z<2.2$ were 
selected using the $UVJ$ diagram and the 3D-HST grism spectra were used to 
unambiguously identify metal absorption lines in the stacked spectra, 
including the G band ($\lambda$4304\AA), MgI ($\lambda$5175\AA), and NaI 
($\lambda$5894\AA), and to spectroscopically derive an average age of 
$1.3^{+0.1}_{-0.3}$~Gyr, in good agreement with our photometrically-derived 
ages in the overlapping redshift interval (see hatched gray box in 
Figure~\ref{fig4}). \citet{whitaker13} also showed that the reddest 80\% of 
the quiescent sample is dominated by metal lines and have older mean stellar 
ages ($1.6^{+0.5}_{-0.4}$ Gyr), whereas the bluest galaxies have strong Balmer 
lines and a spectroscopic age of $0.9^{+0.2}_{-0.1}$~Gyr. This range in stellar 
ages for the population of massive quiescent galaxies at $1.4<z<2.2$ is in 
good agreement with the spread in ages we derived photometrically for the 
sample of quiescent progenitors of UMGs. 

Figure~\ref{fig4} also shows the spectroscopically measured light-weighted ages 
of $z\sim0$ UMGs. 
Specifically, the local measurements plotted in Figure~\ref{fig4} correspond 
to the typical ages of early-type galaxies at $z\sim0$ with 
$\log{(M_{\rm star}/M_{\odot})}\approx 11.8$ in high- and low-density 
environments \citep{thomas05}; the age of early-type galaxies at 
$0.05 < z < 0.06$ with $\log{(M_{\rm star}/M_{\odot})}\approx 11.8$ from the 
Sloan Digital Sky Survey (SDSS; \citealt{thomas10}); the typical age of all 
galaxies at $z\sim0.13$ with $\log{(M_{\rm star}/M_{\odot})}\approx 11.8$ from 
the SDSS \citep{gallazzi05}; and the typical age of early-type galaxies at 
$z\sim0.13$ with $\log{(M_{\rm star}/M_{\odot})}\approx 11.8$ from the SDSS 
\citep{gallazzi06}. 

Our study provides the first direct proof in the early universe of the results 
and implications of the archeological studies of UMGs in the local universe. 
Specifically, from the study of the local stellar population scaling relations 
presented in \citet{gallazzi06}, $z\sim0$ UMGs appear to have a typical 
light-weighted ages of $\sim$10~Gyr, implying a median formation redshift of 
$z \approx 1.9$. The inferred scatter in age at the very high-mass end 
($\sim0.078$-0.088~dex, or $\approx$1.8-2~Gyr; \citealt{gallazzi06}) implies a 
formation redshift for UMGs as low as $z\sim$1.1, and as high as 
$z\sim$3.9-4.2. The formation redshift and spread in age of local UMGs agree 
remarkably well with our findings and our estimated ages for the progenitors 
of UMGs. Indeed, at $2<z<3$, most of the progenitors of $z\sim0$ UMGs are 
star-forming and the massive end of the local quiescent red sequence is being 
vigorously assembled, in agreement with a median formation redshift of UMGs of 
$z\sim2$ from archeology studies. As quiescent progenitors are already present 
at $z=2.75$, their formation must have started at $z>3$ for at least some of 
them, in good agreement with archeological studies in the local universe. 
This is also quantitatively consistent with recent results showing the 
existence of very massive galaxies at $3<z<4$, and the dominance of 
star-bursting galaxies at the massive end (\citealt{marchesini10}; 
\citealt{muzzin13b}). By $1.0<z<1.5$, most star-forming progenitors have 
quenched, in remarkably good agreement with the low formation redshift of 
$z\sim$1.1 from the archeological studies of local UMGs. Moreover, assuming 
that the progenitors of local UMGs start quenching at $2.5<z<3.0$ and that 
their quenching is completed at $1.0<z<1.5$ (since no star-forming progenitors 
are seen at $z<1$), and assuming passive evolution since $z=1$, then the mean 
age difference between the quiescent progenitors of local UMGs should be 
$\sim2.6$~Gyr, i.e., the age difference of the universe between $z=2.75$ and 
$z=1.25$. This is in good agreement with the age dispersion of local UMGs as 
derived from their fossil records (\citealt{gallazzi05}; \citealt{thomas05}; 
\citealt{gallazzi06}; \citealt{thomas10}).

\subsubsection{Star-formation activity}\label{ssfr}

In this section we present a more quantitative analysis of the evolution with 
redshift of the SFRs and the specific star-formation rates 
(sSFR=SFR/$M_{\rm star}$) of the progenitors of UMGs. 

\begin{figure}
\epsscale{1.1}
\plotone{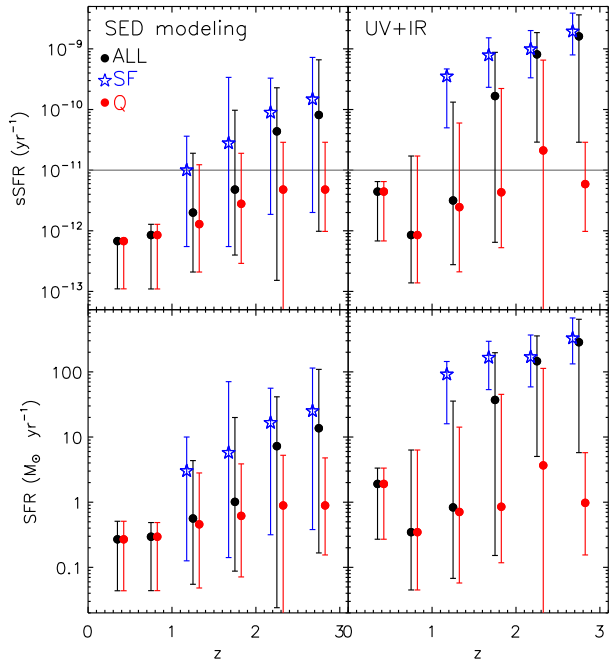}
\caption{Evolution with redshift of the sSFR (top panels) and the SFR (bottom 
panels) derived from SED modeling (left panels) and from the combination of the 
UV and IR luminosities (right panels) for the progenitors of UMGs selected 
with the semi-empirical approach using abundance matching. Symbols as in 
Fig.~\ref{fig4}. The continuous line in the top panels represents the value 
of sSFR$=10^{-11}$~yr$^{-1}$ typically used in the literature to separate 
star-forming from quiescent galaxies.\label{fig5}}
\end{figure}

Figure~\ref{fig5} shows the evolution with redshift of the SFRs and sSFR 
derived from SED modeling (left-hand panels) and from the combination of the 
UV and IR luminosities (right-hand panels) for the progenitors of UMGs.
At $z \lesssim 1.5$, the progenitors of UMGs are predominantly quiescent, with 
sSFR$<10^{\rm -11}$~yr, whereas at $z \gtrsim 1.5$ they become increasingly 
dominated by star-forming galaxies. The median SFRs of the star-forming 
progenitors from the combination of the UV and IR luminosities are 
$\sim$170-330~M$_{\odot}$ yr$^{-1}$ at $1.5<z<3$, i.e., the star-forming 
progenitors are intensely star-bursting galaxies, with 
sSFR$\approx$0.8-2~Gyr$^{-1}$, hence doubling their stellar mass in 
$\sim$0.5-1.2~Gyr if these SFRs were to be sustained over this period of time. 
The evolutions with redshift of the sSFR and SFR derived from SED modeling are 
qualitatively similar to the evolutions derived from the combination of the UV 
and IR luminosities, although they differ quantitatively. Specifically, SFRs 
from the combination of the UV and IR luminosities are typically larger than 
the SFRs from SED modeling, resulting in larger sSFRs. This is especially true 
for the star-forming progenitors, most of which are robustly detected in the 
MIPS 24$\mu$m, with SFRs from the combination of the UV and IR luminosities an 
order of magnitude larger than the SFRs from SED modeling. 

\begin{figure*}
\epsscale{1.}
\plotone{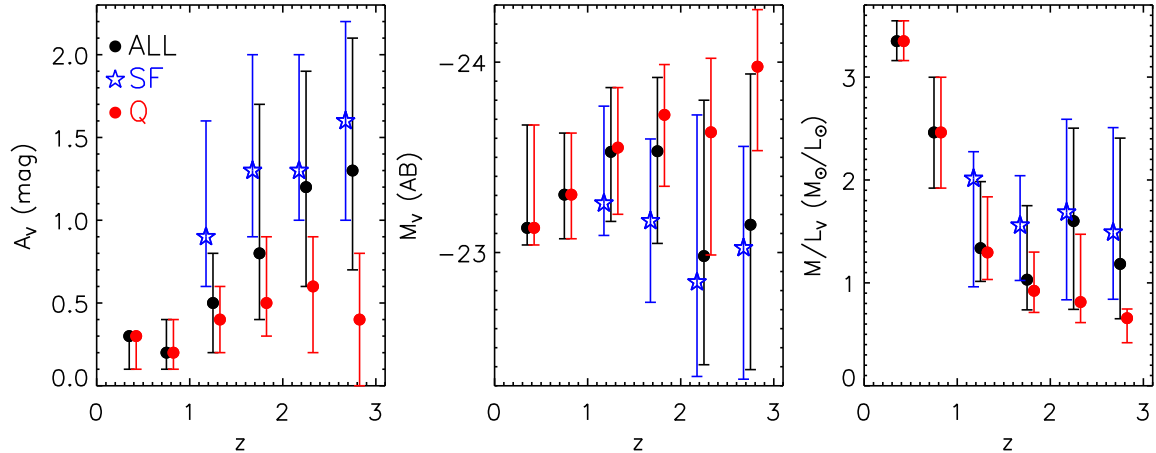}
\caption{Evolution with redshift of the dust extinction (left), the 
rest-frame $V$-band absolute magnitude (middle), and the mass-to-light ratio 
in the rest-frame $V$-band, $M_{\rm star}/L_{\rm V}$, (right) for the 
progenitors of UMGs selected with the semi-empirical approach using abundance 
matching. Symbols as in Fig.~\ref{fig4}. Dust extinction of the 
progenitors increases with redshift, driven by the growing importance of dusty 
star-bursting galaxies among the progenitors at $z>1.5$. A mild increase of 
the dust extinction with redshift in the quiescent progenitors is also 
detected. The $M_{\rm star}/L_{\rm V}$ of the star-forming progenitors remains 
roughly constant with redshift, but larger than the $M_{\rm star}/L_{\rm V}$ of 
the quiescent progenitors at $z>1.5$ due to larger dust obscuration. The 
$M_{\rm star}/L_{\rm V}$ of the quiescent progenitors increases steadily with 
cosmic time due to aging of the overall stellar population.\label{fig6}}
\end{figure*}

It is worth noting that the progenitors selected as quiescent using the 
$UVJ$ diagram become increasingly more actively star forming at high redshift. 
The reason for this is twofold. On one hand, the quiescent galaxies at $z>2$ 
may indeed have non-negligible levels of star-formation activity. As the early 
universe is investigated, the quiescent population will be preferentially 
dominated by post-starburst, relatively younger galaxies whose star-formation 
activity is still being quenched or has only recently been quenched. On the 
other hand, contamination of the $UVJ$ quiescent population due to 
misclassified dusty star-bursting galaxies may increase with redshift and be 
significant at $z>2$. From the $UVJ$ diagram, there is a non-negligible 
fraction of galaxies at $z>2$ straddling the quiescent and star-forming 
regions. Interestingly, most of these galaxies have MIPS 24$\mu$m detections at 
$>5$~$\sigma$, indicative of dusty star-bursting galaxies. The SFRs of 
$\sim$98\% of the quiescent progenitors with robust MIPS detections overlap 
with the low-SFR tail of the distribution of SFRs of the star-forming 
progenitors with robust MIPS detections, suggesting that most of the quiescent 
progenitors with 24$\mu$m detection may be either transitioning objects from 
the star-bursting to the quiescent population, or objects in a phase of 
rejuvenated star formation characterized by relatively low levels of 
star-formation activity.

It should also be kept in mind that the observed MIPS flux could be 
potentially contaminated by MIR emission related to obscured AGNs, whose 
fraction has been shown to increase with stellar mass, especially at higher 
redshifts (e.g., \citealt{kriek07}). While the relative contributions of star 
formation and AGN to the 24$\mu$m flux are difficult to assess here, AGN 
contamination can be partly responsible for the discrepancy between the SFRs 
derived from SED modeling and UV+IR, and it may contribute in making some of 
the $UVJ$ quiescent progenitors appear star-forming in Figure~\ref{fig5}. 
We finally note that low MIPS fluxes can also be fully accounted by 
processes unrelated to obscured AGNs and on-going star formation, such as 
cirrus dust heated by old stellar populations and circumstellar dust embedding 
asymptotic giant branch stars \citep{fumagalli13}.

We therefore caution against taking the values of the SFRs and sSFRs of the 
quiescent galaxies at $z>1.5$ at face value, and argue that these should be 
interpreted as upper limits due to the potential contamination from dusty 
star-bursting galaxies and/or obscured AGNs. Even with these caveats, the 
result that the progenitors of UMGs are increasingly dominated by dusty 
star-bursting galaxies at higher redshifts (i.e., $z>1.5-2$) is robust. 

\subsubsection{Dust extinction}\label{sec-dust}

The left panel of Figure~\ref{fig6} shows the evolution with redshift of the 
dust extinction ($A_{\rm V}$) derived from the modeling of the UV-to-8$\mu$m 
SEDs. The progenitors of UMGs are progressively dustier with increasing 
redshift, with the median dust extinction increasing from 
$A_{\rm V}\sim0.3$~mag at $z=0.35$ to $A_{\rm V}\sim1.3$~mag at $z=2.75$. The 
observed growth is driven by the growing importance with redshift of the 
population of dusty starburst galaxies, having a median extinction of 
$A_{\rm V}\sim1.6$~mag and a 68\% range of $A_{\rm V}\sim1-2.2$~mag at $z=2.75$. 
On the contrary the quiescent population is characterized by dust extinction 
$A_{\rm V} < 1$~mag over the full targeted redshift interval $0.2<z<3.0$.

We note that the quiescent progenitors also show a mild increase with 
redshift of the dust extinction. Evidence for an increase with redshift of the 
dust extinction in both the overall and quiescent populations of massive 
galaxies were previously presented by \citet{whitaker10} and 
\citet{marchesini10}. \citet{whitaker10} found a median extinction of 
$A_{\rm V}=0.2-0.3$~mag in massive ($M_{\rm star}>10^{11}$~M$_{\odot}$) quiescent 
galaxies at $z\sim1-2$, in qualitative agreement with the range of $A_{\rm V}$ 
found in our sample of quiescent progenitors in the overlapping redshift 
range. \citet{marchesini10} studied a complete sample of galaxies at $3<z<4$ 
with $M_{\rm star}>4\times10^{11}$~M$_{\odot}$, finding a median dust extinction 
of $A_{\rm V}=1.4$~mag, in good agreement with the median extinction in 
our sample of progenitors at $2.5<z<3.0$, and evidence of increasing amount of 
dust in the population of very massive galaxies in the early universe.

\subsubsection{Rest-frame absolute $V$-band magnitude}

The central panel of Figure~\ref{fig6} shows the evolution with redshift of the 
rest-frame $V$-band absolute magnitude ($M_{\rm V}$) derived by integrating the 
redshifted rest-frame filter bandpass from the best-fit EAZY template, as 
described in \citet{brammer11}. This method produces similar rest-frame 
luminosities as other methods interpolating between the observed bands that 
bracket the rest-frame band at a given redshift (e.g., \citealt{rudnick03}).

The progenitors of UMGs have median rest-frame $V$-band absolute magnitude in 
the range from $M_{\rm V}=-23.0$ to $M_{\rm V}=-23.5$. When the progenitors are 
separated in quiescent and star-forming galaxies, a clear trend with redshift 
becomes evident. The star-forming progenitors span a large range in 
$M_{\rm V}$ at each redshift (0.7-1.4 mag, depending on the redshift interval). 
The median rest-frame $V$-band absolute magnitude of the star-forming 
progenitors increases with cosmic time from $M_{\rm V}=-22.9$ at $2<z<3$ to 
$M_{\rm V}=-23.3$ at $1.0<z<1.5$, which correspond to 
$1.3\times L^{\star}_{\rm V}$ and 2.5$\times L^{\star}_{\rm V}$, respectively 
(with $L^{\star}_{\rm V}$ the characteristic luminosity of the rest-frame 
$V$-band luminosity function of galaxies at the corresponding redshift; 
\citealt{marchesini12}). On the contrary, the rest-frame $V$-band luminosity of 
the quiescent progenitors decreases with cosmic time from $M_{\rm V}=-24.0$ 
(corresponding to $\sim 3.3 \times L^{\star}_{\rm V}$) at $2.5<z<3.0$ to 
$M_{\rm V}=-23.1$ (corresponding to $\sim 4 \times L^{\star}_{\rm V}$) at 
$0.2<z<0.5$. 

\begin{figure*}
\epsscale{1.}
\plotone{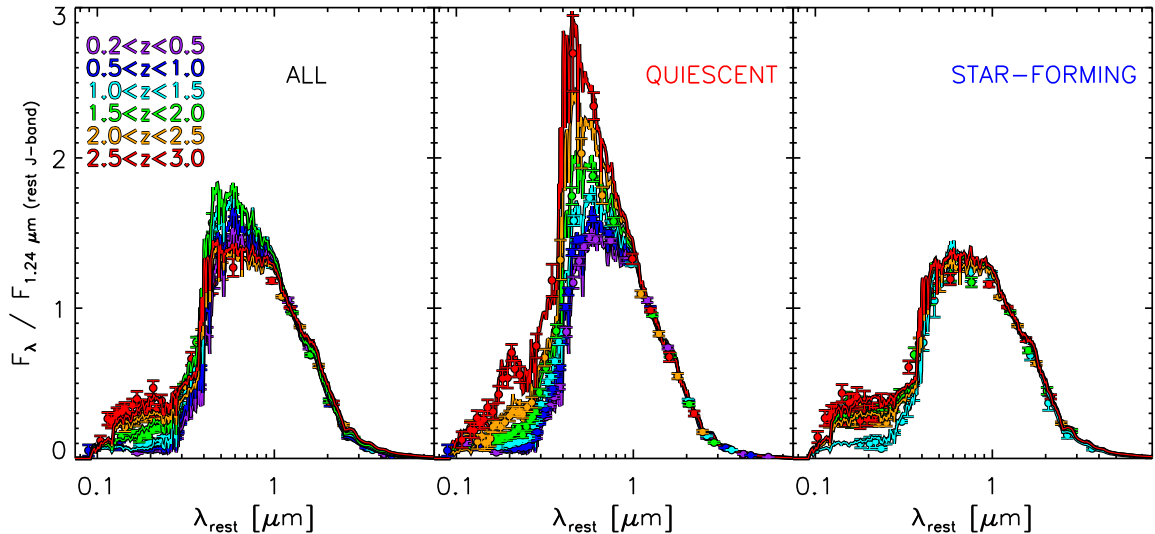}
\caption{Evolution with redshift of the median rest-frame SEDs of the 
progenitors of UMGs selected with the semi-empirical approach using abundance 
matching. The left panel shows all progenitors at different redshifts; the 
middle and right panels show the evolution of the rest-frame SEDs of the 
quiescent and star-forming progenitors, respectively. Filled colored circled 
represent the median rest-frame SEDs derived from the observed photometry, 
whereas the colored curves represent the median best-fits from SED modeling. 
The SEDs derived from these two methods are in good agreement. The 
progenitors of UMGs have rest-frame SEDs typical of dusty star-bursting 
galaxies at high redshift (i.e., $z\gtrsim$2), whereas their rest-frame SEDs 
are dominated at low redshifts by the SEDs typical of quiescent galaxies with 
little or no dust extinction, and old ages. The quiescent progenitors at high 
redshift have rest-frame SEDs typical of post-starburst galaxies, 
characterized by strong Balmer breaks and ``peaky'' rest-frame optical SEDs. 
\label{fig7}}
\end{figure*}

\subsubsection{Mass-to-light ratio}

The right panel of Figure~\ref{fig6} shows the evolution with redshift of the 
mass-to-light ratio $M_{\rm star}/L_{\rm V}$. The $M_{\rm star}/L_{\rm V}$ of the 
progenitors of UMGs is roughly constant at $z>1$ 
($M_{\rm star}/L_{\rm V} \sim 1.1$~M$_{\odot}$~L$^{-1}_{\odot}$) and increases with 
cosmic time to $M_{\rm star}/L_{\rm V} \sim 3.3$~M$_{\odot}$~L$^{-1}_{\odot}$ at 
$z=0.35$. The star-forming and quiescent progenitors show once again different 
dependency of their mass-to-light ratios as a function of redshift. 
The star-forming progenitors are characterized by a constant mass-to-light 
ratio, $M_{\rm star}/L_{\rm V} \sim 1.7$~M$_{\odot}$~L$^{-1}_{\odot}$, over the 
entire $1<z<3$ redshift range. On the contrary, the mass-to-light ratio of the 
quiescent progenitors increases steadily with cosmic time, from 
$M_{\rm star}/L_{\rm V} \sim 0.7$~M$_{\odot}$~L$^{-1}_{\odot}$ at $z=2.75$ to 
$M_{\rm star}/L_{\rm V} \sim 3.3$~M$_{\odot}$~L$^{-1}_{\odot}$ at $z=0.35$. 
Interestingly, at $z>1$, the mass-to-light ratio of the star-forming 
progenitors is larger than the mass-to-light ratio of the quiescent progenitors 
by a factor of $\sim 1.5-2.3$ due to the significantly larger amount of dust 
extinction in the star-forming population. At $z<1$, the mass-to-light ratio 
of the quiescent progenitors increases due to the aging (and the fainting) of 
the overall stellar population.

\subsection{Evolution of the SEDs}\label{seds}

Figure~\ref{fig7} shows the evolution of the rest-frame SEDs of the 
progenitors of UMGs from $z=3$ to $z=0.2$, along with the evolution of the 
rest-frame SEDs of the quiescent and star-forming progenitors (middle and 
right panels, respectively). The SEDs have been normalized using the flux in 
the rest-frame $J$-band (i.e., $\lambda_{\rm rest}=1.24$~$\mu$m). This figure 
summarizes the results for the evolution of the stellar population properties 
found in previous \S\ref{sec-prop}. Whereas little evolution is seen in the 
star-forming progenitors (which dominate at $z>2$), consistently with the 
derived evolution of their stellar population properties, significant 
evolution is observed for the quiescent progenitors, which dominate the 
overall progenitors' population at $z<2$. Specifically, the SEDs of the 
quiescent progenitors evolve from an SED typical of a young post-starburst 
galaxy with moderate dust extinction at $z=2.75$ into an SED typical of a 
quiescent, maximally old stellar population with little or no dust extinction 
at $z=0.35$. As Figure~\ref{fig7} shows, the population of progenitors of UMGs 
was dominated by dusty star-bursting galaxies in the early universe. Such 
population has mostly quenched by $z\sim1.5$, with all progenitors having 
become quiescent by $z=1$, and has aged thereafter with cosmic time, evolving 
into a homogeneous population of quiescent, maximally old progenitors by 
$z=0.35$. 

\subsection{Possible systematics effects}\label{caveats}

In this section we address three possible systematic effects which can 
potentially bias the results presented in the previous sections, namely 
systematic errors in photometric redshift estimates, contamination from 
emission lines, and systematic uncertainties in the modeling of the observed 
SEDs. 

\subsubsection{Systematic errors in photometric redshifts}

Previous searches for old and massive galaxies at $z>4$ highlighted the 
difficulty in unambiguously identifying old and massive objects at extreme 
redshifts on the basis of spectral fitting, with equally acceptable solutions 
between high-redshift very massive galaxies with low-extinction and massive 
($10^{11}$~M$_{\odot}$) galaxies with high extinction ($A_{\rm V}\sim4$~mag) at 
intermediate redshifts ($z\sim2$) (e.g., \citealt{dunlop07}). Moreover, using 
NIR medium band data from the NMBS, \citet{marchesini10} found that the 
inclusion of an ``old and dusty'' template in the photometric redshift 
estimate caused approximately half of the massive galaxies population at $z>3$ 
to be consistent with a somewhat lower photometric redshift in the range 
$2<z<3$. Whereas, at present, there is no strong evidence from spectroscopic 
studies of the existence of such an old and dusty population, 
\citet{muzzin13b} showed that the inclusion of such ``old and dusty'' template 
has significant effects on the high-mass end of the stellar mass function of 
star-forming galaxies at $z>1.5$, especially at $3<z<4$. 

Given that at $z>2$ our sample of progenitors of UMGs are dominated by a 
population of massive galaxies with very red $U-V$ and $V-J$ colors and 
significant amount of dust extinction (see Fig.~\ref{fig8} for four 
representative examples of SEDs), it is worth examining the potential 
systematic effect of including this ``old and dusty'' template on the inferred 
evolution of the progenitors of local UMGs. We have repeated the entire 
analysis after inclusion of the same ``old and dusty'' template adopted in 
\citet{muzzin13b}. Specifically, we used the corresponding stellar mass 
functions from \citet{muzzin13b} to select progenitors and we studied the 
evolution of their properties. 

Consistently with what found by \citet{marchesini10} and \citet{muzzin13b}, we 
find that quiescent galaxies are not affected by the inclusion of this new 
template, whereas the star-forming galaxies with extreme colors tend to 
systematically shift to lower redshifts. Whereas individual star-forming 
galaxies at $z>2$ may change their redshifts substantially, the results agree 
well qualitatively and quantitatively at $z<2.5$, with relevant differences 
only in the highest targeted redshift bin. Specifically, the progenitors at 
$2.5<z<3.0$ tend to be slightly less massive (by $\sim$0.15 dex), less dust 
extincted ($\sim$0.3~mag), and with a slightly bluer median $U-V$ color (by 
$\sim$0.2 mag, although with the same spanned range in $U-V$ color) when the 
``old and dusty'' template is included. Worth noting is also a significant 
increase in the spanned ranges of rest-frame $U-V$ and $V-J$ colors toward 
redder values, especially at $2<z<2.5$, when the new template is included. 

Despite these differences, the overall evolutionary picture of the progenitors 
of local UMGs does not change at all in, e.g., the $U-V$ versus $M_{\rm star}$ 
diagram, hence our main results and conclusions are robust against the 
inclusion of the ``old and dusty'' template, whose relevance has yet to be 
confirmed and assessed spectroscopically.

\subsubsection{Emission line contamination}

Contamination of the photometry from emission lines is another source of 
systematic effects, potentially resulting in significant overestimated stellar 
masses. Several recent spectroscopic investigations have shown serious 
systematic effects due to emission lines on the derived stellar masses of 
rest-frame UV selected, star-forming galaxies at $z>4$ (e.g., 
\citealt{bowler12}; \citealt{ellis13}; \citealt{labbe13}; \citealt{stark13}). 

We have used the best-fit EAZY template to quantify the impact of emission 
lines on the estimated stellar masses. Specifically, the EAZY templates used 
to derive photometric redshifts include emission lines, and we have measured 
emission lines equivalent widths on the best-fit EAZY templates of the 
progenitors' sample. Then, we used the measured equivalent widths to correct 
the observed SEDs, and we re-estimated stellar masses using the emission lines 
corrected SEDs. This modeling leads to conclude that the majority of the 
progenitors' sample at $0.2<z<3.0$ is not affected significantly by emission 
line contamination, with a mean difference of 0.03$\pm$0.05~dex in the stellar 
mass estimated before and after emission line correction. We estimate that 
$\sim$5\% of the sample appears to be affected significantly by emission line 
contamination, with stellar masses overestimated on average by a factor of 
$\sim$2.5. Since these galaxies represent less than 10\% of the sample of 
star-forming progenitors at $z>1.5$, we conclude that, based on the emission 
line ratios and simplistic modeling adopted in EAZY, our results appears to be 
relatively robust against emission line contamination. 

However, we caution that unusually high line ratios, especially 
[OIII]/H$\beta$, are being consistently found in star-forming galaxies at 
$z>3$, significantly biasing estimated stellar masses \citep{holden14}. While 
these studies have been currently targeting only rest-frame UV selected 
galaxies with smaller stellar masses ($\log{(M_{\rm star}/M_{\odot})}<10.7$) 
compared to our progenitors' sample, we stress that a robust quantitative 
assessment of emission line contamination in the $z>2$ galaxy population at 
the high-mass end will necessarily require spectroscopic studies for large 
samples of massive star-forming galaxies, currently missing.

\subsubsection{SED modeling}\label{s-sedmodel}

The derived stellar population properties depend on the adopted choice of 
SED-modeling assumptions. We have investigated the effects of different 
SED-modeling assumptions by adopting different SFHs, extinction curves, and 
metallicities. Specifically, in place of the exponentially-declining SFH we 
have used a delayed-$\tau$ SFH of the form $SFR \propto t~exp(-t/\tau)$, which 
allows for increasing SFR at earlier times. In place of the \citet{calzetti00} 
extinction curve, we have adopted both the Milky Way extinction law 
\citep{cardelli89} and the extinction curve proposed by \citet{kriek13}, which 
assumes a dust attenuation curve with $R_{\rm V}=4.05$ and a UV dust bump 
which is 20\% of the strength of the Milky Way bump, following the 
prescription by \citet{noll09}. Finally, we have relaxed the assumption on 
metallicity by leaving it as a free parameter in the SED modeling. 

The systematic effects on the stellar population properties of the progenitor 
galaxies are found to be significantly smaller than the corresponding typical 
random uncertainties for most of the different SED-modeling assumptions. For 
the stellar mass, the systematic effect is found to be relevant only when 
adopting the Milky Way extinction curve, although still comparable to the 
typical random uncertainty, with $M_{\rm star}$ smaller by 0.05(0.09) dex for 
the quiescent (star-forming) progenitors. The adoption of the Milky Way 
extinction curve also results in the largest systematic effects for the dust 
extinction, although significantly smaller compared to its typical random 
uncertainty, with $A_{\rm V}$ smaller by 0.05(0.2)~mag for the quiescent 
(star-forming) progenitors compared to the case when the \citet{calzetti00} 
law is adopted. The largest effect on the star-formation rate for the 
quiescent progenitors is found when the delayed-$\tau$ model is adopted, with 
SFR smaller by 0.1~dex, whereas the largest effect on the star-formation rate 
for the star-forming progenitors is found when the Milky Way extinction curve 
is adopted, with SFR smaller by 0.13~dex. Finally, the largest effects on the 
stellar ages of the star-forming progenitors are found when the delayed-$\tau$ 
model is adopted, with $\langle t \rangle_{\rm SFR}$ larger by $\sim$0.1~dex 
compared to the default SED-modeling assumptions. The stellar ages of the 
quiescent progenitors changes at most by 0.02~dex depending on the different 
SED-modeling assumptions. We therefore conclude that the inferred evolution 
with redshift of the stellar population properties of the quiescent and 
star-forming progenitors of local UMGs are robust and not very sensitive to 
reasonable choices of the SED-modeling assumptions.

However, systematic effects in the modeling of the observed SEDs, and hence in 
the estimated stellar masses, can be relevant in galaxies characterized by 
composite stellar populations. Specifically, the stellar masses of the dusty, 
massive star-forming progenitors at $z>1.5$ could be potentially overestimated 
when modeled with a single stellar population, especially if characterized by 
multiple stellar population components with differential dust extinctions, 
such as a quiescent un-obscured population and a young, dusty star-bursting 
component. 

To investigate potential biases in the stellar masses of the star-forming 
progenitors at $2<z<3$, we have derived stellar masses by allowing for 
multiple stellar components to contribute to the observed SEDs of the dusty 
star-forming progenitors. No systematic differences are found, with a median 
difference smaller than 0.1~dex and a scatter of $\sim$0.5~dex (Brammer et al. 
2014, in preparation). We also note that the observed SEDs of most of the 
dusty star-forming progenitors of UMGs at $z>1.5$ cannot be well modeled by an 
ad hoc combination of a quiescent un-obscured population and a young, dusty 
star-bursting component, which would arguably cause the largest biases in the 
stellar mass estimates (Brammer et al. 2014, in preparation). 

We therefore conclude that, while the stellar masses of individual 
star-forming progenitors may be systematically different by a factor of a 
few-to-several due to simplified SED-modeling assumptions, the stellar mass 
distribution of the overall population appears to be largely unaffected by the 
adopted SED-modeling assumptions. Nonetheless, we stress that the 
multi-component modeling in Brammer et al. (2014) is still over simplified 
(e.g., a dust screen with an adopted \citealt{calzetti00} extinction curve), 
and systematic errors up to a factor of several in stellar mass cannot be 
excluded for a subset of the most extreme star-forming progenitors. 

\subsubsection{The reddest star-forming progenitors: a new population?}

\begin{figure}
\epsscale{1.15}
\plotone{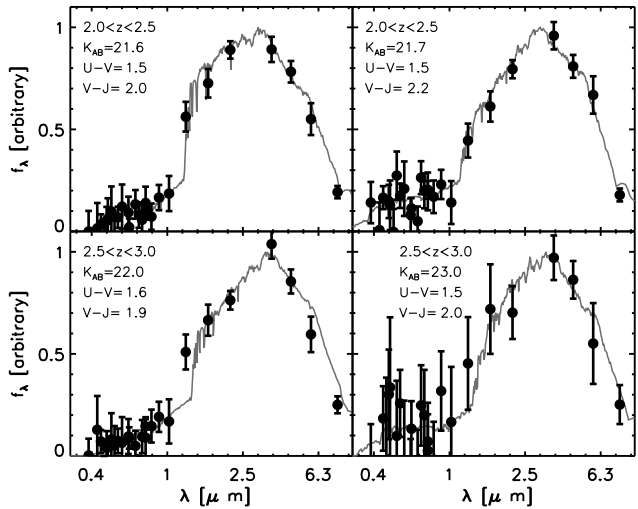}
\caption{Examples of SEDs of reddest star-forming progenitors with $V-J>1.9$. 
Black filled circles and error bars represent the observed photometry, whereas 
the continuous gray curve shows the best-fit model. The top and bottom rows 
show SEDs at $2.0<z<2.5$ and $2.5<z<3.0$, respectively. \label{fig8}}
\end{figure}

In this study we have demonstrated that a significant fraction of the 
progenitors of UMGs at $z>2$ are dusty star-forming galaxies. Amongst that 
population, $\sim$19\% have extraordinary SEDs with very red $V-J$ colors 
($V-J>1.9$).  As shown in Figure~\ref{fig2}, effectively zero galaxies  with 
such red $V-J$ colors exist at $z<1$ (down to the UltraVISTA mass limit), and 
therefore it is tempting to classify them as a ``new'' population of galaxies.

These extremely red galaxies are intriguing, and in Figure~\ref{fig8} we show 
four representative examples of their SEDs. The solid lines are the best-fit 
FAST SEDs using the default SED modeling assumptions.  The fits are reasonable 
and hence with the default assumptions we naturally interpret this population 
as massive, extremely red, star-forming progenitors.  However, the SEDs are 
quite unlike local galaxies, and given that they only appear at $z>1$, it is 
important to consider carefully what else this unusual population might be and 
how accurate their stellar masses are.  As already discussed, systematic 
effects likely dominate the uncertainties in both the photometric redshifts 
and SED modeling of individual galaxies among this extreme population. 

Specifically, strong emission lines (or unusual emission-line ratios) could 
result in significantly overestimated stellar masses, while the presence of 
optically thick regions could even cause the underestimating of their stellar 
masses. Perhaps of most concern are composite stellar populations and/or 
potential blending/confusion of sources in ground-based images which could 
result in the unusual SEDs and therefore biased stellar masses. 

Without additional data it is difficult to say how reasonable the SED modeling 
assumptions are for this population, and how reliable their stellar masses are. 
Therefore, currently we advise caution in interpreting this extreme population. 
While they are only $\sim$19\% of the overall population, and unlikely to 
change our overall picture of UMG formation, they are intriguing, and detailed 
investigations, beyond the capabilities of current photometric catalogs, will 
be needed to confirm the existence of this population, and to assess its 
relevance in the formation of massive galaxies in the early universe. 

We note that a better understanding of this population could be made with 
current instrumentation. Spectroscopy from ground-based near-infrared 
spectrographs and ALMA could measure spectroscopic redshifts and clearly 
identify the level of contamination of the broad band colors from emission 
lines (if any).  Likewise, {\it Hubble Space Telescope} near-infrared 
observations could resolve them spatially and be used to test for 
confusion/blending, as well as potential systematics from resolved composite 
stellar populations. Lastly, deep spectroscopy could be used to determine 
velocity dispersions, which will be the key for understanding how reliable the 
photometrically-derived masses are. Followup of this extreme population will 
be key for a more complete understanding of the evolutionary history of UMGs.

\section{Summary and Conclusions}\label{sec-discussion}

In this paper we have investigated the evolution since $z=3$ of the properties 
of the progenitors of ultra-massive galaxies 
($\log{(M_{\rm star}/M_{\odot})}\approx11.8$) at $z=0$ (UMGs). The progenitors 
have been selected using the fixed cumulative number density approach and a 
semi-empirical method using abundance matching in the $\Lambda$CDM paradigm. 
When the progenitors are selected with the fixed cumulative number density 
approach, the stellar mass of the progenitors is seen to grow by a factor of 
$\sim2.2$ from $z=3$ to $z=0$, implying that about half of the stellar mass in 
local UMGs was assembled at $z>3$. If the progenitors are selected with the 
semi-empirical method using abundance matching, the evolution in stellar mass 
is much more significant ($0.56^{+0.35}_{-0.25}$~dex), with only about a fourth 
of the stellar mass in local UMGs having been assembled by $z=3$. Since the 
semi-empirical approach using abundance matching accounts for merging, whereas 
the fixed cumulative number density method does not, the former is a more 
desirable and robust approach to constrain the evolution of the median 
cumulative number density and therefore to select progenitors. The comparison 
between the assembly history of local brightest cluster galaxies (BCGs) as 
predicted from semi-analytic models with the inferred evolution in stellar 
mass since $z=3$ of the progenitors of local UMGs provides yet another 
manifestation of the previously highlighted tensions between the observed and 
model-predicted evolutions of the stellar mass function since $z=4$ (e.g., 
\citealt{marchesini09}). Specifically, low-mass galaxies in the stellar mass 
range $10^{9}-10^{11}$~M$_{\odot}$ form too early in the models and are too 
passive at later times \citep{fontanot09}, with the mass growth of simulated 
BCGs being dominated by accretion since $z\sim4$ \citep{delucia07}, in 
contrast to what is measured in our study. 

The progenitors of UMGs have been separated in quiescent and star-forming 
galaxies using the rest-frame $U-V$ versus $V-J$ diagram ($UVJ$ diagram). At 
$z=0.35$, the progenitors of UMGs constitute a homogeneous population 
with similar rest-frame $U-V$ and $V-J$ colors, typical of quiescent 
galaxies with old stellar populations (age$\sim$7~Gyr). As the progenitors are 
followed to higher redshifts, this population becomes increasingly 
diversified. At $z<1$, the progenitors are all quiescent, while at $z>1$ the 
contribution from star-forming galaxies progressively increases, with the 
progenitors' population being dominated by dusty, star-bursting galaxies at 
$2<z<3$. 

The stellar population properties (i.e., stellar age, dust extinction, 
star-formation rate, specific star-formation rate, rest-frame absolute 
$V$-band magnitude, and mass-to-light ratio in the rest-frame $V$ band) have 
been estimated from the modeling of the UV-to-8$\mu$m spectral energy 
distributions and 24$\mu$m data, and their evolutions with redshift have been 
investigated. At $z=2.75$, most of the progenitors of UMGs are dusty 
($A_{\rm V}\sim$1-2.2~mag) star-bursting (SFR$\sim$100-700~M$_{\odot}$~yr$^{-1}$; 
sSFR$\sim$0.8-4~Gyr$^{-1}$) galaxies with a large range in stellar ages, i.e., 
0.3-1.6~Gyr. At the same redshift, the quiescent progenitors have properties 
typical of young (age $\sim$0.6-1 Gyr) post-starburst galaxies with little 
dust extinction ($A_{\rm V}\sim$0.4 mag) and strong Balmer breaks, showing a 
large scatter in rest-frame $U-V$ color ($\sim$0.2 mag). With decreasing 
redshift, the fraction of quiescent progenitors increases as the star-forming 
progenitors quench to build the quiescent population. At the same time, the 
stellar population of the already quiescent progenitors is clearly seen to age 
with cosmic time, with their properties changing accordingly. Specifically, 
their stellar ages and mass-to-light ratio increase with decreasing redshift, 
and their median SEDs evolve from being characterized by a strong Balmer break 
to being shaped by the 4000\AA~break in the rest-frame optical. Moreover, the 
quiescent progenitors are seen to migrate from the bottom-left to the 
upper-right corner of the $UVJ$ quiescent box and their rest-frame $U-V$ color 
to redden, as expected from an aging quiescent stellar population. 

It is interesting to note that the stellar properties of the star-forming 
progenitors do not appear to be changing substantially over the redshift 
interval $1.5<z<3.0$, and that over this same interval the scatter in the 
$U-V$ color of the quiescent progenitors remains constant and large 
($\sim$0.17~mag) and it starts to decrease only at $z<1.5$. It appears indeed 
that $z=1.5$ marks the time by which most of the star-forming progenitors have 
quenched, as indicated by the very few star-forming progenitors at 
$1.0<z<1.5$. Consistent with this interpretation, the last star-forming 
progenitors to quench at $z=1.25$ are characterized by lower levels of star 
formation and older ages on average compared to the star-forming progenitors 
at $z>1.5$. By $z=1$, all star-forming progenitors have quenched, and the 
quiescent progenitors are simply left aging thereafter in the following 
$\sim$7.7~Gyr to $z=0$, and some additional growing due to, e.g., minor dry 
mergers. 

Whereas the existence at $z=2.75$ of a small fraction ($\sim$15\%) of 
quiescent progenitors implies that the assembly of the very massive end of the 
local quiescent red-sequence population must have started at $z>3$, our 
results suggest it had been mostly assembled between $z=3$ and $z=1.5$, with 
all the growth by in-situ star formation ceased by $z=1$. Most of the 
quenching of the star-forming progenitors is seen to happen between $z=2.75$ 
and $z=1.25$, implying a mean age difference of $\sim$2.6~Gyr among the UMGs. 
Our results are in remarkably good agreement with the typical formation 
redshift of quiescent, red-sequence UMGs in the local universe ($z\sim1.9$) 
and their rms scatter in age ($\approx$1.8-2~Gyr) as derived from 
archeological studies (e.g., \citealt{gallazzi06}), and provide a direct probe 
in the early universe of the results and implications of the fossil records in 
the population of UMGs today.

\begin{figure}
\epsscale{1.1}
\plotone{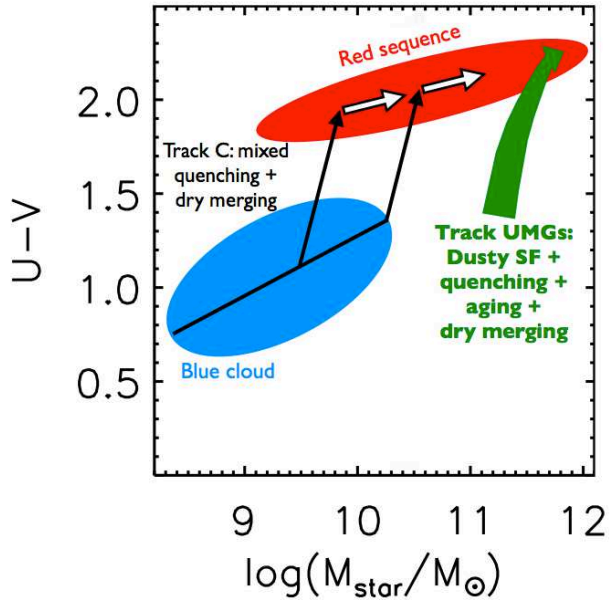}
\caption{Schematic evolutionary tracks plotted in the color-mass diagram 
showing possible paths for the formation of massive red-sequence galaxies in 
the local universe. Track C, proposed by \citet{faber07}, is a ``mixed'' 
scenario involving both early and late quenching of blue, star-forming 
galaxies. The red sequence galaxy arise from the blue galaxy when star 
formation is quenched (upward black arrows from the blue cloud to the red 
sequence). Once the galaxy arrives on the red sequence, it may evolve more 
slowly along it through a series of dry mergers (white arrows). The green 
arrow represents our newly proposed path for the formation of UMGs in the 
local universe based on our findings. This new scenario involves in-situ early 
stellar growth during a short and intense dusty burst of star formation at 
which point the progenitor manifest as a red, massive, heavily dust-extincted 
star-bursting galaxy. After quenching, the progenitor continues reddening due 
to aging and potentially growing in mass via dry mergers, reaching its final 
position on the color-mass diagram. In this new scenario, the progenitor of 
local UMGs have never lived on the blue star-forming cloud since $z=3$. The 
population of red galaxies is seen to change with cosmic time, dominated by 
dusty star-bursting galaxies at $z>2$ and by quiescent galaxies at lower 
redshift. \label{fig9}}
\end{figure}

Finally, we have shown that the progenitors of UMGs have never lived on the 
blue star-forming cloud in the last 11.4~Gyr of cosmic history, i.e., since 
$z=3$. This is true also for the star-forming progenitors, which have a median 
rest-frame $U-V$ color at $z=2.75$ similar to the quiescent progenitors, with 
a significant fraction of them having $U-V$ colors much redder than most of 
the quiescent progenitors at the same redshift. In light of our findings, 
the picture in which early-type UMGs at $z\sim0$ have formed from star-forming 
galaxies living on the blue cloud that reddened after quenching and further 
grew through moderate dry merging has to be refined. Such scenario is pictured 
in Figure~\ref{fig9} as taken from the ``Evolution Track C'' in 
\citet{faber07}. This scenario involves early mass assembly and star formation 
in galaxies on the blue cloud of star-forming galaxies. When the 
star-formation activity in the blue galaxies gets quenched, they rapidly move 
onto the red sequence. Once on the red sequence of quiescent galaxies, any 
further growth in stellar mass in these progenitors mostly proceeds via dry 
merging among galaxies on the red sequence. Whereas such path is still 
viable to explain the formation of less massive (e.g., $10^{11}$~M$_{\odot}$) 
local red-sequence galaxies which may have gone through mostly unobscured 
(hence blue) episodes of early star formation, our study suggest an 
alternative path for the formation of UMGs at $z\sim0$. This new scenario, 
pictured as a green arrow in Figure~\ref{fig9}, involves early mass assembly 
and stellar growth in a short and intense dusty burst of star formation during 
which the progenitors manifest as a red, heavily dust-obscured star-bursting 
galaxies. After quenching, the progenitors continue reddening due to the aging 
of their stellar populations, and eventually grow in mass through (mostly 
minor) dry merging, finally reaching their current positions on the color-mass 
diagram. This scenario qualitatively agrees with the predictions from the 
theoretical model of the co-evolution of massive galaxies and super-massive 
black holes originally proposed in \citet{granato04} and further developed in 
\citet{cai13} and \citet{lapi14}, which indicates that the star formation in 
the progenitors of local very massive spheroids proceeded within a heavily 
dust-enshrouded medium at $z>1.5$ over a timescale $\lesssim 0.5-1$~Gyr.

Our study provides a complete and consistent picture of how the most massive 
galaxies in the local universe have formed and assembled since $z=3$. While 
we see the star-forming progenitors disappearing and turning into the quiescent 
progenitors, we do not know what physical mechanisms are responsible for 
halting the star formation. Is the AGN, if present, playing a major role? What 
about feedback from the starburst itself? For the first time, our analysis 
provides a sample of star-forming progenitors that must quench by $z=1$. 
Detailed studies of the AGN content, physical conditions of the gas (e.g., gas 
fraction and gas-phase metallicity), stellar population properties (e.g., 
stellar ages and metallicities, SFRs), structural and dynamical properties, 
and environment of these star-forming progenitors can shed light into the 
physical processes responsible for the inevitable quenching of their 
star-formation activity. High $S/N$ optical and near-infrared spectroscopic 
studies, as well as deep sub-millimeter studies with ALMA will provide the 
datasets to answer the remaining outstanding questions. 

Needless to say, our study needs to be extended to earlier times to 
investigate the properties of the progenitors at $z>3$, i.e., in the first 2 
billion years of cosmic history. Have the progenitors ever lived on the blue 
star-forming cloud? Which are the progenitors of the quiescent, massive, 
post-starburst galaxies we see at $2.5<z<3$? Much deeper near-infrared 
selected catalogs over a large surveyed area are required to push this 
investigation to higher redshift. The second data release (DR2) of the 
UltraVISTA survey will provide much deeper near-infrared coverage over 
$\sim$70\% of the full COSMOS field, reaching $\sim1.2$~mag deeper in the 
$K_{\rm S}$ band, and allowing to probe the stellar population 0.5~dex deeper 
in stellar mass. 

The observed assembly since $z=3$ of the most massive galaxies need 
to be investigated in more detail in the context of theoretical models of 
galaxy formation and evolution. The new generation of semi-analytic models 
accounting for a better treatment of satellite galaxies, for the formation of 
the intra-cluster component, including detailed chemical enrichment models, 
and coupled with radiative transfer codes or templates to predict the infrared 
luminosities of model galaxies (e.g., \citealt{vega05}; \citealt{lagos12}) 
will provide valuable tools to further test the physical mechanisms driving 
the formation and evolution of today's most massive galaxies. 

Finally, we note that $\sim$19\% of the star-forming progenitors at $z>2$ have 
extraordinary SEDs with very red $V-J$ colors ($V-J>1.9$), potentially 
representing a ``new'' population of galaxies in the early universe that 
effectively disappear by $z\sim1$. While unlikely to change our overall 
picture of the formation of UMGs, systematic effects dominate the 
uncertainties in both the photometric redshifts and SED modeling of individual 
galaxies among this extreme population. Detailed investigations, including 
spectroscopy with near-infrared spectrographs and ALMA, and near-infrared 
imaging with the {\it Hubble Space Telescope}, will be needed to confirm the 
existence of this intriguing population and for a more complete understanding 
of the evolutionary history of UMGs.


\acknowledgments
We are grateful to the anonymous referee whose comments and suggestions helped 
improving significantly this paper. We gratefully acknowledge Peter Behroozi 
for providing the code implementing the semi-empirical approach using 
abundance matching to select progenitors. DM acknowledges the support of the 
Research Corporation for Science Advancement's Cottrell Scholarship. DM also 
acknowledges support from Tufts University Mellon Research Fellowship in Arts 
and Sciences. AM and MF acknowledge support from the Netherlands Foundation 
for Research (NWO) Spinoza grant. MS acknowledges support from the programs 
HST-GO-12286.11 and HST-GO-12060.95, provided by NASA through a grant from the 
Space Telescope Science Institute, which is operated by the Association of 
Universities for Research in Astronomy, Incorporated, under the NASA contract 
NAS5-26555. BMJ and JPUF acknowledge support from the ERC-StG grant 
EGGS-278202. JSD acknowledges the support of the European Research Council 
through the award of an Advanced Grant, and the support of the Royal Society 
via a Wolfson Research Merit Award. JSD also acknowledges the contribution of 
the EC FP7 SPACE project ASTRODEEP (Ref.No: 312725). The Dark Cosmology Centre 
is funded by the DNRF. This study is based on data products from observations 
made with ESO Telescopes at the La Silla Paranal Observatory under ESO 
programme ID 179.A-2005 and on data products produced by TERAPIX and the 
Cambridge Astronomy Survey Unit on behalf of the UltraVISTA consortium.


\appendix

\section{Evolution of the Properties of Progenitors of UMGs Selected Using 
the Fixed Cumulative Number Density Method}\label{appendixA}

\begin{figure*}
\epsscale{1.1}
\plotone{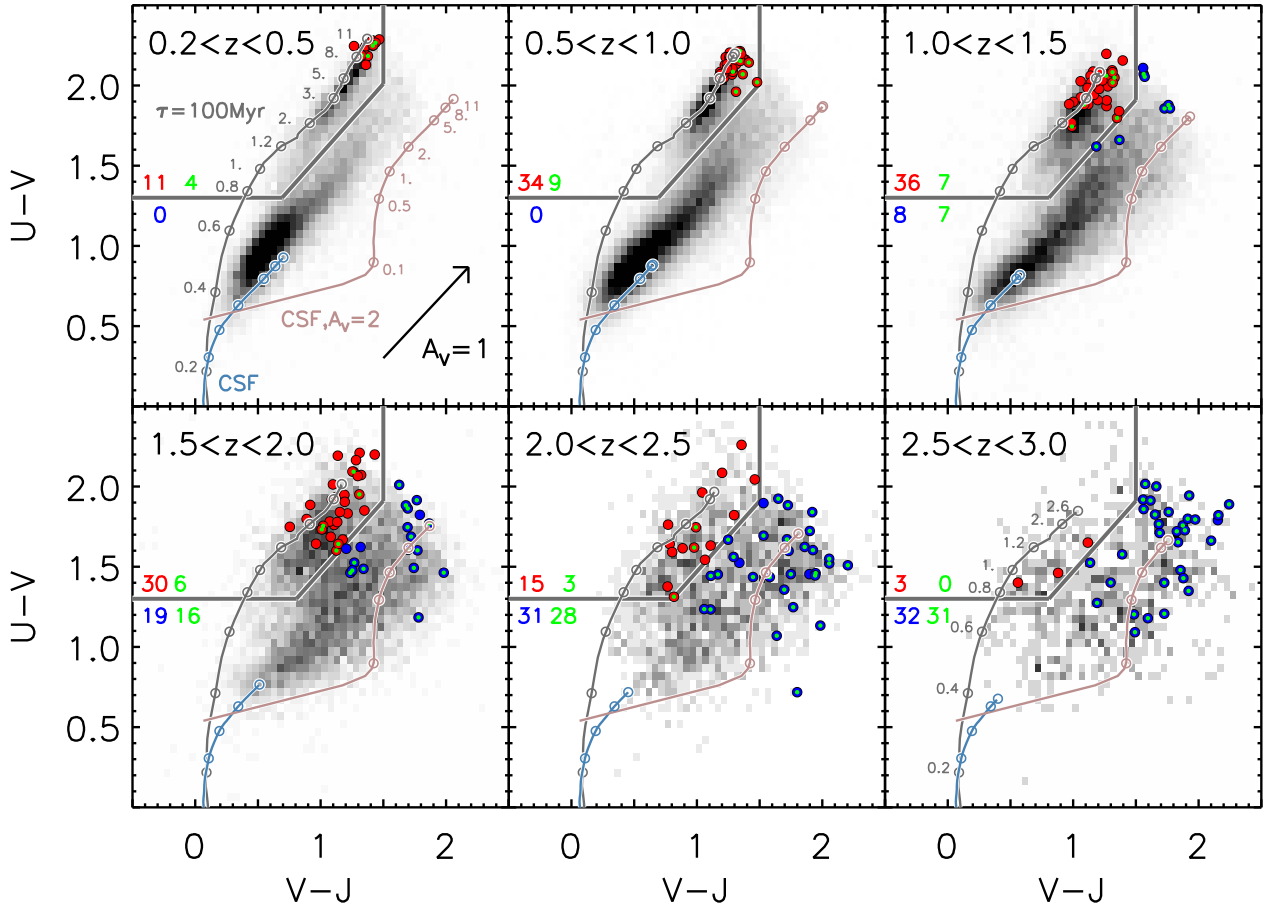}
\caption{Rest-frame $U-V$ versus $V-J$ diagram in the targeted redshift 
intervals between $z=0.2$ and $z=3.0$ for the progenitors of local UMGs 
selected with the fixed cumulative number density approach. Symbols as in 
Figure~\ref{fig2}.\label{fig2app}}
\end{figure*}

\begin{figure*}
\epsscale{1.1}
\plotone{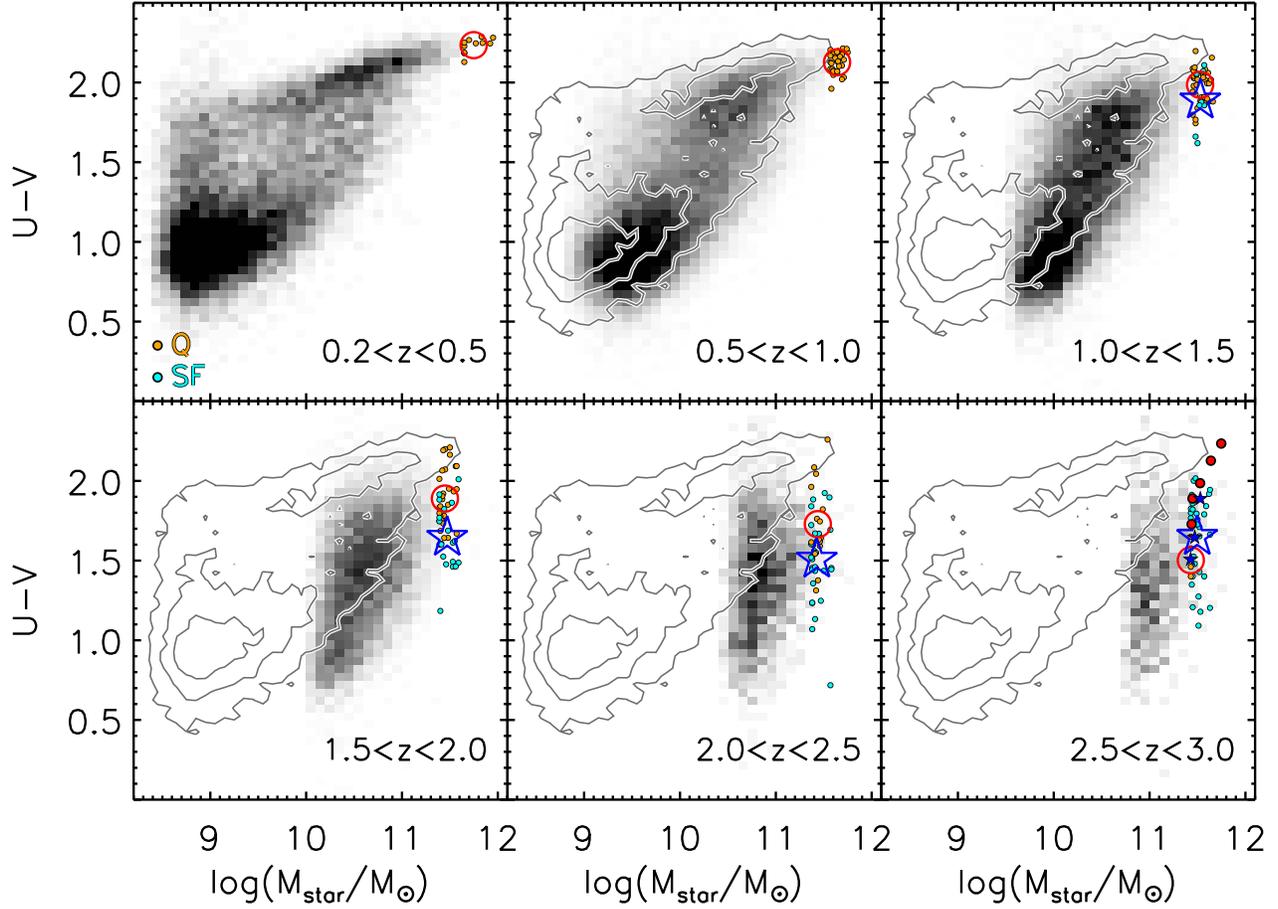}
\caption{Rest-frame $U-V$ color versus stellar mass diagram for the 
progenitors of local UMGs selected with the fixed cumulative number density 
approach. Symbols as in Figure~\ref{fig3}.\label{fig3app}}
\end{figure*}

\begin{figure}
\epsscale{0.6}
\plotone{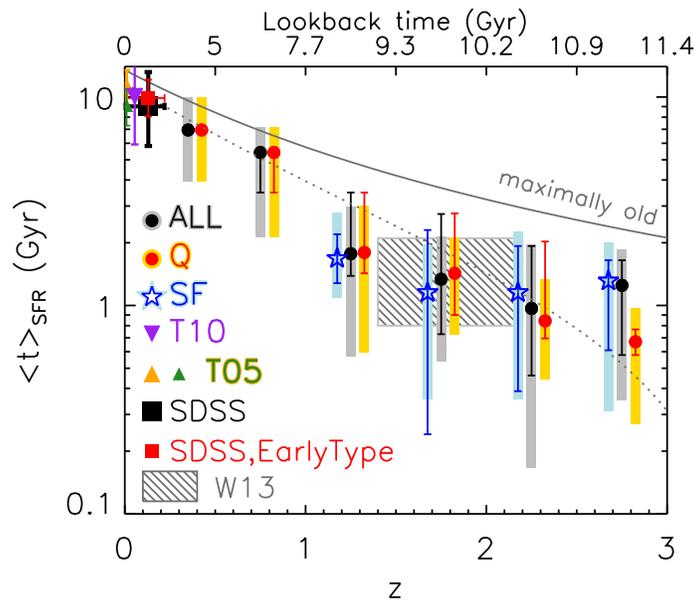}
\caption{Evolution with redshift of the stellar age 
$\langle t \rangle_{\rm SFR}$ derived from the modeling of the UV-to-8$\mu$m 
SEDs of the progenitors of local UMGs selected with the fixed cumulative 
number density approach. Symbols as in Figure~\ref{fig4}.\label{fig4app}}
\end{figure}

\begin{figure}
\epsscale{0.6}
\plotone{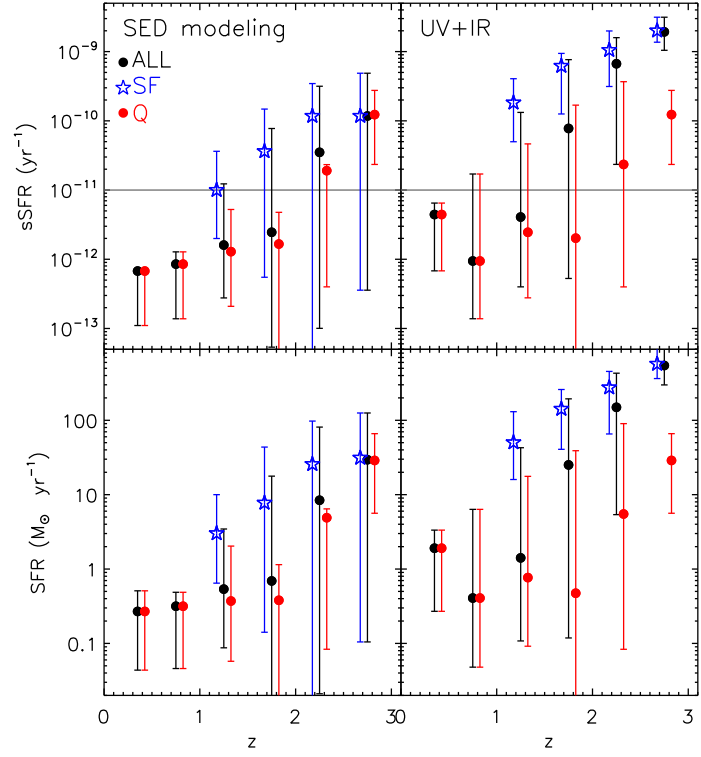}
\caption{Evolution with redshift of the sSFR (top panels) and the SFR (bottom 
panels) derived from SED modeling (left panels) and from the combination of the 
UV and IR luminosities (right panels) for the progenitors of UMGs selected 
with the fixed cumulative number density approach. Symbols as in 
Figure~\ref{fig5}.\label{fig5app}}
\end{figure}

\begin{figure*}
\epsscale{1.}
\plotone{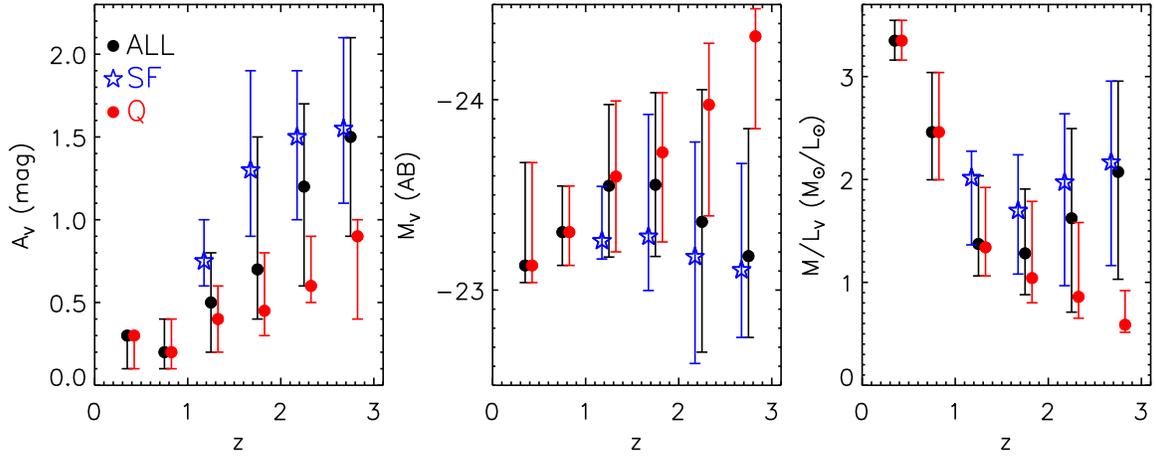}
\caption{Evolution with redshift of the dust extinction (left), the 
rest-frame $V$-band absolute magnitude (middle), and the mass-to-light ratio 
in the rest-frame $V$-band, $M_{\rm star}/L_{\rm V}$, (right) for the 
progenitors of UMGs selected with the fixed cumulative number density 
approach. Symbols as in Figure~\ref{fig6}.\label{fig6app}}
\end{figure*}

\begin{figure*}
\epsscale{1.}
\plotone{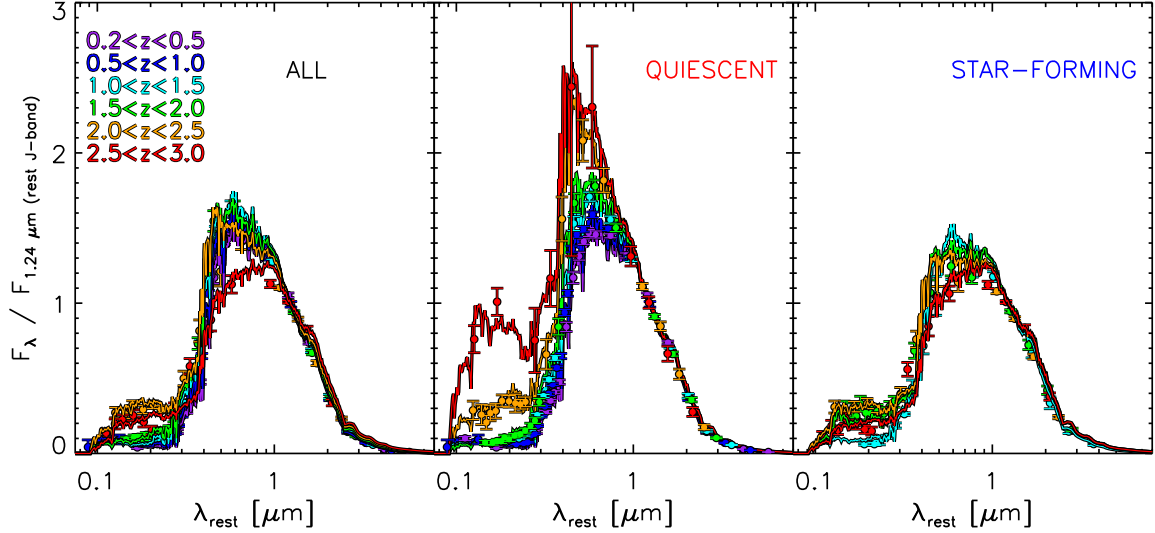}
\caption{Evolution with redshift of the median rest-frame SEDs of the 
progenitors of UMGs selected with the fixed cumulative number density 
approach. Symbols as in Figure~\ref{fig7}.\label{fig7app}}
\end{figure*}

Here we present the evolution of the properties of the progenitors of local 
UMGs selected using the fixed cumulative number density method. 
Figures~\ref{fig2app} and \ref{fig3app} show the evolution of the $UVJ$ 
diagram and the $U-V$ versus $M_{\rm star}$ diagram, respectively. 
Figures~\ref{fig4app}, \ref{fig5app}, and \ref{fig6app} show the evolution as 
a function of redshift of the stellar age, SFR and sSFR, and dust extinction, 
rest-frame $V$-band magnitude, and mass-to-light ratio, respectively. Finally, 
Figure~\ref{fig7app} shows the evolution of the SEDs as a function of 
redshift. As shown by these figures, the evolution in the stellar population 
parameters (e.g., rest-frame colors, level of star 
formation activity, stellar ages, extinction, rest-frame optical absolute 
magnitudes, and mass-to-light ratios) inferred when the progenitors are 
selected with the fixed cumulative number density method is qualitatively 
similar and quantitatively consistent to the evolution obtained when the 
progenitors are selected adopting the semi-empirical approach using abundance 
matching.

\end{document}